\documentclass[hyper]{prop2015}

\usepackage{amsfonts,amsmath,amsthm}
\usepackage{mathrsfs}
\usepackage{bbm}
\usepackage{enumitem}
\usepackage[all]{xy}
\usepackage{tikz}
\usetikzlibrary{arrows}
\usetikzlibrary{patterns}
\usetikzlibrary{decorations.markings}

\renewcommand{\hom}{\mathsf{hom}}
\newcommand{\sfAbGrp}{\mathsf{AbGrp}}
\newcommand{\sfAlg}{\mathsf{Alg}}
\newcommand{\sfB}{\mathsf{B}}
\newcommand{\sfd}{\mathsf{d}}
\newcommand{\eps}{\varepsilon}

\newcommand{\sff}{\mathsf{f}}

\newcommand{\sfG}{\mathsf{G}}
\newcommand{\sfGoodAlg}{\mathsf{GoodAlg}}
\newcommand{\sfGrp}{\mathsf{Grp}}
\newcommand{\sfhom}{\mathsf{hom}}

\newcommand{\sfj}{\mathsf{j}}

\newcommand{\sfMaps}{\mathsf{Maps}}
\newcommand{\sfMfd}{\mathsf{Mfd}}
\newcommand{\sfScheme}{\mathsf{Scheme}}
\newcommand{\CN}{\mathcal{N}}
\newcommand{\sfSet}{\mathsf{Set}}
\newcommand{\sfSMfd}{\mathsf{SMfd}}
\newcommand{\sfSpac}{\mathsf{Spc}}
\newcommand{\sfTop}{\mathsf{Top}}

\newcommand{\sU}{\mathsf{U}}     			
\newcommand{\sfVect}{\mathsf{Vect}}
\newcommand{\CL}{\mathcal{L}}

\newcommand{\scrC}{\mathscr{C}}
\newcommand{\scrD}{\mathscr{D}}

\newcommand{\IN}{\mathbbm{N}}
\newcommand{\IK}{\mathbbm{K}}
\newcommand{\IR}{\mathbbm{R}}
\newcommand{\IZ}{\mathbbm{Z}}

\newcommand{\id}{{\rm id}}
\newcommand{\im}{{\rm im}}
\newcommand{\dd}{{\rm d}}
\newcommand{\myxymatrix}[1]{\vcenter{\vbox{\xymatrix{#1}}}}

\theoremstyle{plain}
\newtheorem{theorem}{Theorem}[section]

\theoremstyle{definition}

\newtheorem{example}[theorem]{Example}

\category{Proceedings}


\title{Higher Structures in M-Theory}
\subtitle{\href{http://www.maths.dur.ac.uk/lms/109/index.html}{LMS/EPSRC Durham Symposium on Higher Structures in M-Theory}}

\author[B. Jur\v{c}o]{Branislav Jur{\v c}o\inst{a}}
\author[C. Saemann]{Christian S{\"a}mann\inst{b}}
\author[U. Schreiber]{Urs Schreiber\inst{c}}
\author[M. Wolf]{Martin Wolf\inst{d}\footnote{Corresponding author e-mail:~\href{m.wolf@surrey.ac.uk}{\textsf{m.wolf@surrey.ac.uk}}}}

\address[1]{Charles University, Faculty of Mathematics and Physics, Mathematical Institute, Prague 186 75, Czech Republic}
\address[2]{Maxwell Institute for Mathematical Sciences, Department of Mathematics, Heriot--Watt University, Edinburgh EH14 4AS, United Kingdom; EMPG--19--06}
\address[3]{Division of Science and Mathematics, New York University, Abu Dhabi, United Arab Emirates, on leave from Czech Academy of Science}
\address[4]{Department of Mathematics, University of Surrey, Guildford GU2 7XH, United Kingdom; DMUS--MP--19--02}

\shortauthors{B.~Jur{\v c}o, C.~S{\"a}mann, U.~Schreiber, M.~Wolf}

\begin{abstract}
The key open problem of string theory remains its non-perturbative completion to M-theory. A decisive hint to its inner workings comes from  numerous appearances of \emph{higher structures} in the limits of M-theory that are already understood, such as higher degree flux fields and their dualities, or the higher algebraic structures governing closed string field theory. These are all controlled by the higher homotopy theory of derived categories, generalised cohomology theories, and  $L_\infty$-algebras. This is the introductory chapter to the proceedings of the LMS/EPSRC Durham Symposium on `Higher Structures in M-Theory':
\begin{center}
\includegraphics[width=.9\textwidth]{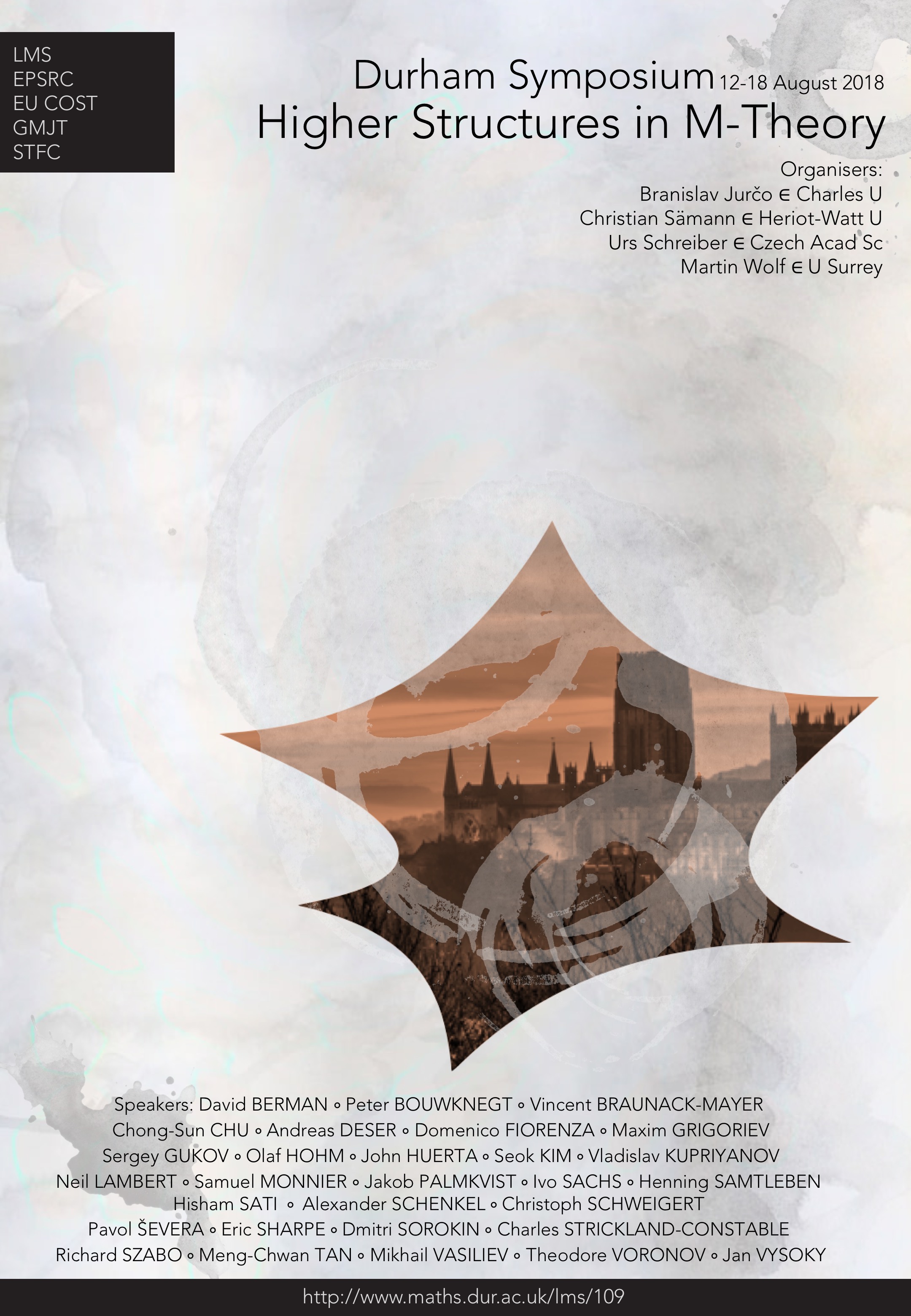}
\end{center}
We first review higher structures as well as their motivation in string theory and beyond. Then we list the contributions in this volume, putting them into context. 
\end{abstract}

\begin{acknowledgements}
We would like to express our gratitude to the \href{https://www.lms.ac.uk/}{London Mathematical Society} and the \href{https://epsrc.ukri.org/}{Engineering and Physical Sciences Research Council} for providing the majority of the funding for the conference. We are also very grateful to the \href{http://www.cost.eu/COST_Actions/mpns/MP1405}{EU COST Action MP1405 Quantum Structure of Space-Time} and the \href{https://stfc.ukri.org/}{Science and Technology Facilities Council} for significant contributions to the travel budget of this conference. Moreover, we acknowledge the kind contribution of the \href{http://www.icms.org.uk/activities/gmjt.php}{Glasgow Mathematical Journal Trust Learning and Research Support Fund} to the travel cost of the Scottish participants. We would also like to thank the local organisers at Durham University, especially Dirk Sch{\"u}tz, who all contributed vitally to making the symposium a success. Finally, we thank all the participants of the symposium for creating an inspiring atmosphere and for many interesting discussions.
\end{acknowledgements}

\shortabstract

\begin{document}

\maketitle

\section{Contributors}

The following authors have contributed to these conference proceedings:
I.~Bandos,
M.~Benini, 
D.~S.~Berman,
V.~Braunack-Mayer, 
S.~Bruinsma, 
C.-S.~Chu,
A.~Deser, 
F.~Farakos,
D.~Fiorenza, 
J.~Fuchs,
M.~Grigoriev,
O.~Hohm,
J.~Huerta, 
B.~Jur{\v c}o,
A.~Kotov,
V.~G.~Kupriyanov, 
N.~Lambert,
S.~Lanza, 
T.~Macrelli,
L.~Martucci,
S.~Monnier,
L.~Raspollini, 
I.~Sachs, 
C.~S{\"a}mann,
H.~Samtleben,
H.~Sati, 
A.~Schenkel,
L.~Schmidt,
U.~Schreiber,
C.~Schweigert, 
E.~Sharpe,
D.~Sorokin, 
C.~Strickland-Constable,
R.~J.~Szabo,
Th.~Th.~Voronov,
J.~Vysok\'y,
M.~Wolf, and
R.~Zucchini.

\section{Introduction}

\begin{table*}
\centering
\scalebox{0.98}{
\begin{proptabular}[1cm]{c|c}{The basic higher structures in string theory are visible directly in the first quantised string. They are essentially related to the fact that the charged spinning string is a higher-dimensional analogue of the charged spinning particle.}
\label{tab:BasicHigherStructures}
Ordinary Structure (Particle) & Higher Structure (String) \\
gauge potential 1-form $A$ & Kalb--Ramond 2-form $B$\\
gauge bundle for Dirac charge quantisation & principal 2-bundle (or gerbe) for Freed--Witten anomaly\\
Lie algebra symmetry $\leftrightarrow$ first-order BRST ghost fields & Lie 2-algebra (or 2-term $L_\infty$-algebra) symmetry $\leftrightarrow$ second-order BRST ghost fields\\
spin bundle $\leftrightarrow$ spin structure (second Whitehead stage) & string 2-bundle for Green--Schwarz anomaly $\leftrightarrow$ string structure (third Whitehead stage)\\
$\vdots$ & $\vdots$
\end{proptabular}
}
\end{table*}

During the twenty years since its inception, the conjecture of M-theory~\cite{Witten:1998uk}, see also~\cite{Duff:1999baa}, has greatly contributed to the understanding of string theory, and its basic ideas have found their way into the textbooks, see e.g.~\cite{Becker:2007zj}. Nevertheless, a proper definition or rather a formulation of M-theory as a coherent theory, and, consequently, a formulation of a full non-perturbative completion of string theory, remains an open problem to date~\cite[Section 12]{Moore:2014aaa}.

This lack of a proper formulation of non-perturbative string theory is problematic as is witnessed, for instance, in recent debates about the existence or non-existence of de Sitter vacua in string theory~\cite{Danielsson:2018ztv,Obied:2018sgi,Akrami:2018ylq}.

A key issue in formulating M-theory is that its underlying principles have remained unclear. If available hints are anything to go by, M-theory is not simply going to be defined by a Lagrangian, nor a scattering matrix, nor any other traditional structure of quantum physics. The suggestions that M-theory might be defined indirectly by plain quantum field theory, either as matrix model quantum mechanics or via the AdS/CFT correspondence, interesting as they are, all suffer from requiring certain limits and backgrounds that seem to preclude a formulation of the full theory.

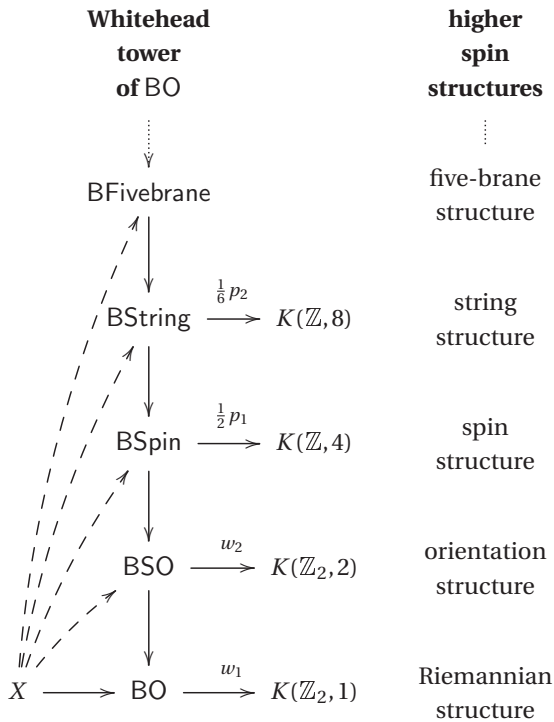
\begin{figure}[h]
\vspace{-.5cm}
\begin{equation*}
\kern.3cm\xymatrix@C=9pt@R=9pt{
& \mbox{$\begin{array}{c}{\bf Whitehead}\\{\bf tower}\\ {\bf of~\mathsf{BO}}\end{array}$}  \ar@{..>}[d] & & \mbox{$\begin{array}{c}{\bf higher}\\{\bf spin}\\ {\bf structures}\end{array}$} \ar@{..}[d]\\
&\mathsf{BFivebrane} \ar[d] & & \mbox{$\begin{array}{c}\text{five-brane}\\{\rm structure}\end{array}$}\\
&\mathsf{BString}  \ar[d] \ar[r]^{\frac16p_2}& K(\IZ,8)& \mbox{$\begin{array}{c}{\rm string}\\{\rm structure}\end{array}$}\\
&\mathsf{BSpin}  \ar[d] \ar[r]^{\frac12p_1}& K(\IZ,4)& \mbox{$\begin{array}{c}{\rm spin}\\{\rm structure}\end{array}$}\\
&\mathsf{BSO}  \ar[d] \ar[r]^{w_2}& K(\IZ_2,2)& \mbox{$\begin{array}{c}{\rm orientation}\\{\rm structure}\end{array}$}\\
X\ar[r]\ar@/^.3pc/@{-->}[ur]\ar@/^.4pc/@{-->}[uur]\ar@/^.6pc/@{-->}[uuur]\ar@/^.8pc/@{-->}[uuuur]&\mathsf{BO} \ar[r]^{w_1}& K(\IZ_2,1)& \mbox{$\begin{array}{c}{\rm Riemannian}\\{\rm structure}\end{array}$}
}
\end{equation*}
\vspace{-1.1cm}
\caption{Shown is the tower of classifying spaces $\mathsf{BG}$ (see Example \ref{exa:ClassifyingSpaces}) of higher analogs of the group $\sfG = \mathsf{Spin}$. Where the latter is obtained from the special orthogonal group $\mathsf{SO}$ by universally cancelling the first homotopy group, the \emph{2-group} (see Example \ref{exa:2Groups}) $\mathsf{String}$ is obtained by universally cancelling the next higher non-trivial homotopy group, namely the third. Each `hook' in the figure is a homotopy fibre sequence which means that the cohomology classes shown measure the obstructions to equipping space-time with the given structure. The obstruction $\tfrac{1}{2}p_1$ to string structure, hence the higher analog to the obstruction $w_2$ to spin structure, embodies the Green--Schwarz anomaly cancellation for the heterotic string \cite{Bunke:2009cba, Sati:2009ic}. The next higher fractional classes $p_2$ appear in the anomaly polynomial of the heterotic five-brane \cite{Sati:2008kz} (see also \cite{Sati:2014}). The spaces $K$ are Eilenberg--MacLane spaces (see Example \ref{exa:DoubleDimRed}). Furthermore, the $p_i$ denotes the $i$-th Pontryagin class and $w_i$ the $i$-th Stiefel--Whitney class of the tangent bundle $TX$ of $X$.
}
\label{fig:WhiteheadTower}
\end{figure}

However, one set of ideas that has been at the forefront from the early days of string theory is the appearance of what nowadays is encapsulated under the notion of \emph{higher structures}. This is essentially short for \emph{higher homotopy structures} (as in higher homotopy groups). Such structures manifest themselves in the higher degrees that are carried by string theoretic objects as compared to their field theoretic counterparts, notably the higher degrees of the flux fields, and the resulting higher order gauge-of-gauge transformations. See Table~\ref{tab:BasicHigherStructures} for some details. Generally, the appearance of homotopy theory is simply the consequence of the \emph{gauge principle} in physics.

Indeed, higher structures are already seen in basic first quantised bosonic string theory. The coupling of the bosonic string to the Kalb--Ramond 2-form $B$-field is the higher degree 2 analogue of the coupling of the charged particle to the 1-form vector potential, and the $B$-field \emph{gerbe} which governs the corresponding Freed--Witten anomaly cancellation \cite{Freed:1999vc,Carey:548736} is nothing but the higher principal \emph{2-bundle} version \cite{Nikolaus:1207ab} of the line bundle which implements Dirac charge quantisation for the charged particle. Moreover, for the first quantised heterotic string one observes that the (twisted) \emph{string structure}, which mathematically reflects its Green--Schwarz anomaly cancellation \cite{Bunke:2009cba,Sati:2009ic}, is the higher analogue of spin structures for spinning particles, in the direct sense of appearing one step higher in the homotopy Whitehead tower of the orthogonal group (see Figure \ref{fig:WhiteheadTower}).

Once we switch to the `second quantisation' of the string known as string field theory, higher algebraic structures become ubiquitous. In particular, the Hilbert space of closed string field theory carries a higher algebraic structure known as an $L_\infty$-algebra~\cite{Zwiebach:1992ie}. While bosonic open string field theory has a Chern--Simons like formulation~\cite{Witten:1985cc}, its supersymmetric completion also requires the introduction of a higher homotopy algebra known as an $A_\infty$-algebra, see e.g.~\cite{Erler:2013xta}.

Given the fundamental role of string theory, and therefore string field theory, is supposed to play in physics, it is not surprising that higher structures appear in many different contexts within (mathematical) physics such as:

\smallskip
\begin{enumerate}[label=\roman*),leftmargin=*]
\item {\bf Field theory:}
\begin{enumerate}[label=\alph*),leftmargin=*]
\item Any classical field theory gives rise to a (quantum) homotopy algebra, a higher algebraic structure governing its dynamics by means of the BRST/BV formalism;
\item The AKSZ construction of field theories~\cite{Alexandrov:1995kv} is based on symplectic higher Lie algebroids;
\item The abstract definition of a topological quantum field theory makes use of higher categories of cobordisms, see e.g.~\cite{Baez:1995:6073-6105};
\item Many moduli spaces of field theories are best described using stacks, which are essentially higher versions of manifolds;
\item The six-dimensional superconformal field theories based on the $\CN=(1,0)$ or $\CN=(2,0)$ tensor multiplet contain a 2-form gauge potential which is part of a connection on a higher principal bundle;
\end{enumerate}
\smallskip
\item{\bf Supergravity:}
\begin{enumerate}[label=\alph*),leftmargin=*]
\item Higher-degree differential form fields that appear in supergravity belong to connections on higher principal bundles;
\item The tensor hierarchies appearing in gauged supergravity are (twisted) higher gauge theories;
\end{enumerate}
\pagebreak
\item {\bf Models of (quantum) space-time:}
\begin{enumerate}[label=\alph*),leftmargin=*]
\item T-duality suggests a generalisation of modelling spacetime by ordinary manifolds. This leads to orbifolds and, more generally, stacks.
\item For quantum space-times, non-commutativity is only first step. Non-associativity follows from higher geometric quantisation, and ultimately the algebra of smooth functions is replaced by an $A_\infty$-algebra. The fuzzy funnel of M2-branes ending on M5-branes shows the necessity for such a higher quantisation in M-theory~\cite{Basu:2004ed}.
\end{enumerate}
\smallskip
\item {\bf Mathematical physics:}
\begin{enumerate}[label=\alph*),leftmargin=*]
 \item The transition from point particles to strings implies the transition from a space-time manifold to loop space. Bundles over loop space correspond to higher bundles over the space-time manifold: an Abelian gerbe over a manifold $M$ maps to an Abelian principal bundle over the loop space of $M$ and a spin structure on $\CL M$ corresponds to a string structure on $M$;
 \item The Courant algebroids of generalised geometry, which form a key mathematical structure underlying T-duality, are best seen as higher symplectic Lie algebroids~\cite{Roytenberg:2002nu}.
 \end{enumerate}
\end{enumerate}

The above mentioned T-duality, which interchanges the momentum modes of a string with its winding modes around a compact cycle, is part of a huge \emph{web of dualities}, linking various string and M-theory vacuum configurations. Much of the fascination with string theory is based on this structure and improving our understanding of string theory dualities is essential if progress is to be made with defining M-theory. T-duality, in particular, is interesting since it sets apart string theory from ordinary particle (quantum) field theory.  Another important example is \emph{S-duality}, a generalisation of electromagnetic duality, which links strongly coupled field theories to weakly coupled ones. Understanding S-duality is certainly one of the aims of the intense study of the superconformal six-dimensional field theory known as the (2,0)-theory: compactifying this theory in different ways to four dimensions yields an important example of S-duality. As we shall explain further below, the mathematical language underlying dualities is category theory, and it is encouraging that this fits nicely with the above mentioned higher homotopical structures arising from the gauge principle.

Altogether, we arrive at the correspondences in Table~\ref{tab:meaning} and we shall explore these in the main part of this article.
\begin{center}
\begin{proptabular}[1cm]{c|c|c}{The meaning of higher/categorical structures.}
\label{tab:meaning}
&&\\[-10pt]
category theory & is really & the theory of duality \\
\begin{tabular}{c}$\infty$-category theory\\ $=$ homotopy theory \end{tabular} & is really & the gauge principle\\
\begin{tabular}{c} $\infty$-topos theory\\ $=$ higher geometry \end{tabular}
 & is really & \begin{tabular}{c}the geometry seen by\\ classical sigma models \end{tabular}
\end{proptabular}
\end{center}

\section{The idea of higher structures}\label{sec:ideas}

\subsection{Basics of category theory}

In the following, we concisely summarise some basic facts on category theory. For a detailed account on category theory and its applications we recommend the text books \cite{Lawvere:1997baa, Leinster:1612.09375, Riehl17}.

\begin{figure}[h]
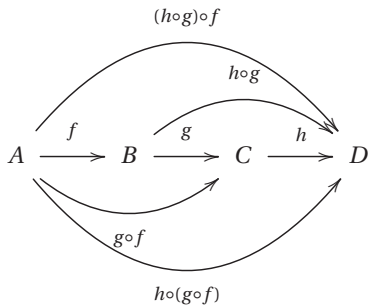

\vspace{-1cm}
\begin{equation*}
\kern1.5cm\myxymatrix{
A \ar@/^10ex/[rrr]^{(h\circ g)\circ f} \ar@/_10ex/[rrr]_{h\circ (g\circ f)} \ar@/_5ex/[rr]_{g\circ f} \ar[r]^{f}
& B \ar@/^5ex/[rr]^{h\circ g} \ar[r]^{g}
& C \ar[r]^{h}
& D
}
\end{equation*}
\vspace{-1cm}
\caption{A category is a \emph{relational set}: in addition to the collection of elements or \emph{objects} $A,B,C,\ldots$, a category provides a relation amongst these in terms of structure-preserving maps $f,g,h,\ldots$ called \emph{morphisms}. Morphisms compose associatively whenever defined and as shown.}
\label{fig:ACategory}
\vspace{-.5cm}
\end{figure}

Mathematical objects such as sets, groups, vector spaces, etc.\ are always studied together with corresponding \emph{(homo)morphisms} or structure preserving functions. The mathematical gadget combining both objects and morphisms is a \emph{category}. A category can thus be seen as a \emph{relational set}: a collection of objects, with information how these may be mapped to each other in a structure preserving way, see Figure~\ref{fig:ACategory}.

Categories are, of course, completely elementary. They could be
introduced in elementary school in place of set theory Venn diagrams~\cite{Lawvere:1997baa}. This is contrary to the fortunately less and less widely spread belief that category theory is esoteric.

Categories are as simple as they turn out to be profound and useful. Category theory proves that the nature of the objects in any category may entirely be reconstructed from the system of morphisms relating them. This means that category theory provides for mathematics something like what physicists value as \emph{Mach's principle} and what philosophers call \emph{structuralism} \cite{AWODEY:1996:209-237}---hence just what we need for understanding dualities and higher structures in string theory.

The basic examples of categories include the \emph{concrete categories} listed in Table~\ref{tab:ConcreteCategories} whose objects are sets equipped with mathematical structure in the classical sense of Bourbaki (for instance, sets equipped with a group structure or with a topology), and whose morphisms are functions between these sets that respect that structure (for instance, group homomorphisms or continuous functions, respectively).

We often write $\mathsf{Obj}(\scrC)$ and $\mathsf{Mor}(\scrC)$ for the objects and morphisms of a category, respectively. The collection of morphisms from an object $C_1\in\mathsf{Obj}(\scrC)$ to another object $C_2\in\mathsf{Obj}(\scrC)$ will be denoted by $\sfhom_\scrC(C_1,C_2)$.

\begin{center}
\begin{proptabular}[1cm]{c|c|c}{The evident examples of categories have as objects mathematical structures in the form of sets with certain operations and properties on them, and as morphisms structure-preserving maps. Already these basic examples are sufficient and necessary for proving statements such as Gelfand duality, which secretly underlie much of modern physics.}
\label{tab:ConcreteCategories}
Category & Objects & Morphisms \\
$\sfSet$ & sets & functions\\
$\sfTop$ & topological spaces & continuous functions\\
$\sfMfd$ & smooth manifolds & smooth functions\\
$\sfVect$ & vector spaces & linear functions\\
$\sfGrp$ & groups & group homomorphisms\\
$\sfAlg$ & algebras & algebra homomorphisms\\
$\vdots$ & $\vdots$ & $\vdots$
\end{proptabular}
\end{center}

\begin{figure}[h]
\vspace{-1cm}
\begin{equation*}
\xymatrix@L=5pt@C=38pt@R=15pt{
& C_2 \ar[dl]_{f_2}  \ar@{|->}[dddr]^F &&\\
C_3 \ar@{|->}[dddr]^F & & \ar[ll]^{f_2\circ f_1}C_1 \ar[ul]_{f_1} \ar@{|->}[dddr]^F&\\
&&&\\
&& F(C_2) \ar[dl]_{F(f_2)}  &\\
&F(C_3) & & \ar[ul]_{F(f_1)} F(C_1) \ar[ll]^{F(f_2\circ f_1)} \\
}
\end{equation*}
\vspace{-1cm}
\caption{A functor $F:\scrC\to\scrD$ between two categories $\scrC$ and $\scrD$ is a map that preserves the relations between the objects.}
\label{fig:IllustrationFunctor}
\vspace{-.5cm}
\end{figure}
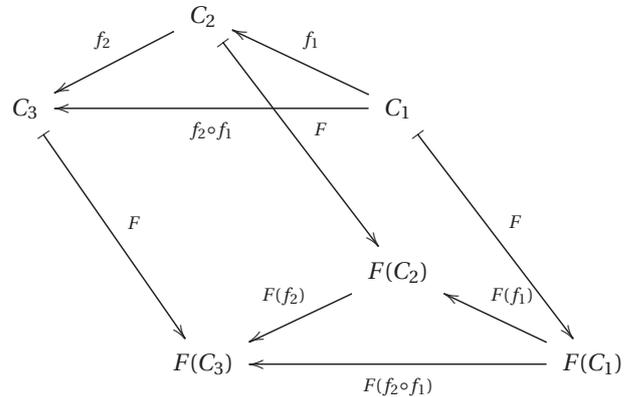

Having introduced categories as relational sets, we would clearly also like to have a notion of morphism between categories encoding relations between them. Such a morphism is called \emph{functor}, and it is simply a pair of maps between the objects and the morphisms of two categories which preserve the relations between the objects that are encoded in the morphisms, see Figure~\ref{fig:IllustrationFunctor}. Put together, small categories\footnote{That is, categories with objects and morphisms forming sets. This restriction avoids the analogue of Russell's paradox for categories.} and functors form themselves the category of small categories $\mathsf{Cat}$.

Examples of functors are abundant, and here we shall only give some basic examples as well as two examples in the context of string theory. Many more examples will make their appearance later on.

\smallskip
\begin{example}
{\bf (Opposite categories and functors)} For every category $\scrC$, we may define what is known as the \emph{opposite category} $\scrC^{\text{op}}$ which is simply the category that has the same objects as $\scrC$ and with the morphisms of $\scrC$ but `going the other way around', that is, the directions of all morphisms are reversed. A classical example of a functor is the \emph{dual vector space functor} $\sfVect^{\text{op}}\to\sfVect$ which assigns to each vector space its vector space dual. Another important example is the functor $\sfTop^{\text{op}}\to\sfAbGrp$ from the opposite category of topological spaces to the category of Abelian groups. This is simply the categorical definition of a \emph{presheaf}. In addition, for every functor $F:\scrC\to\scrD$ there is an \emph{opposite functor} $F^{\text{op}}:\scrD^{\text{op}}\to\scrC^{\text{op}}$.
\end{example}

\begin{figure}[h]
\vspace{-.4cm}
\begin{eqnarray*}
&&\kern1.7cm\myxymatrix{
\sfTop \ar[rr]^{(-)\times S^1}_{\substack{\text{Cartesian}\\[1pt] \text{product functor}}} && \sfTop
}\\
&&\substack{p\text{-brane}\\ \text{embedding}\\ \text{field}}\left\{\xymatrixcolsep{5.2pc}\myxymatrix{
\Sigma \ar[d]_{\phi} \ar@{|->}[r] & \Sigma\times S^1\ar[d]_{\phi\times\id_{S^1}}\\
X \ar@{|->}[r] & X\times S^1
}\right\}\substack{\text{wrapped}\\ (p+1)\text{-brane}\\ \text{embedding}\\ \text{field}}
\end{eqnarray*}
\vspace{-.7cm}
\caption{The functor that constructs wrapped brane configurations. Specifically, the assignment that sends any topological space $X$ to the product space $X \times S^1$ extends to a functor on the category $\sfTop$ of topological spaces. If one thinks of a given continuous function $\phi:\Sigma\to X$ as a $p$-brane embedding field for a brane world volume $\Sigma$ into a space-time $X$, then its image $\phi\times\id_{S^1}:\Sigma \times S^1\to X \times S^1$ under this functor is the embedding field of the corresponding wrapped $(p+1)$-brane, that is, the $p$-brane with one extra compact direction.}
\label{fig:WrappedBraneFunctor}
\vspace{-.4cm}
\end{figure}
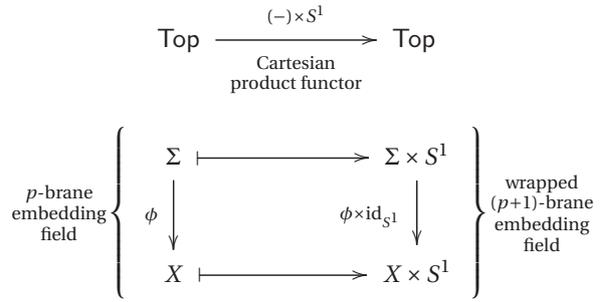

\smallskip
\begin{example}\label{ex:wrapped_branes}
{\bf\mathversion{bold} (Compact dimensions and wrapped $p$-branes)}
In every category in which for any two objects $A$ and $B$ there exists also their \emph{Cartesian product} $A\times B$ (for instance topological product spaces or direct product groups) the assignment $A \mapsto A \times B$ extends to a functor from the category to itself in an evident way. If the ambient category is that of smooth manifolds, or at least that of topological spaces, and if $Y = S^1$ is the circle, then we may think of $X \times S^1$ as the space-time obtained from any given space-time $X$ by adding a single compact dimension. The functoriality of this assignment then encodes the \emph{wrapping} of $p$-branes around this extra dimension, see Figure~\ref{fig:WrappedBraneFunctor}.
\end{example}

\begin{figure}[h]
\vspace{-.4cm}
\begin{eqnarray*}
&&\kern1.6cm\myxymatrix{
\sfTop \ar[rr]^{\sfMaps(S^1,-)}_{\substack{\text{free}\\[1pt] \text{loop space functor}}} && \sfTop
}\\
&&\kern-.2cm\substack{\text{Kaluza--Klein}\\[1pt] \text{reduction}}\left\{\xymatrixcolsep{3.8pc}\myxymatrix{
X \ar[d]_{p} \ar@{|->}[r] & \sfMaps(S^1,X)\ar[d]_{\sfMaps(S^1,p)}\\
Y \ar@{|->}[r] & \sfMaps(S^1,Y)
}\right\}\substack{\text{forget}\\ \text{closed string}\\ \text{configurations}\\ \text{in Kaluza--Klein}\\[2pt] \text{fibres}}
\end{eqnarray*}
\vspace{-.7cm}
\caption{The functor that Kaluza--Klein compactifies closed string configurations. Specifically, the assignment that sends any topological space $X$ to its free loop space $\sfMaps(S^1,X)$ extends to the category $\sfTop$ of topological spaces. If one thinks of a given continuous function $p:X\to Y$ as a Kaluza--Klein compactification of a higher-dimensional space-time $X$ to a lower-dimensional space-time $Y$, then its image $\sfMaps(S^1,p):\sfMaps(S^1,X)\to\sfMaps(S^1,Y)$ under this functor projects out the components of closed string configurations that disappear under the Kaluza--Klein compactification.}
\label{fig:FreeLoopSpaceFunctor}
\vspace{-.4cm}
\end{figure}
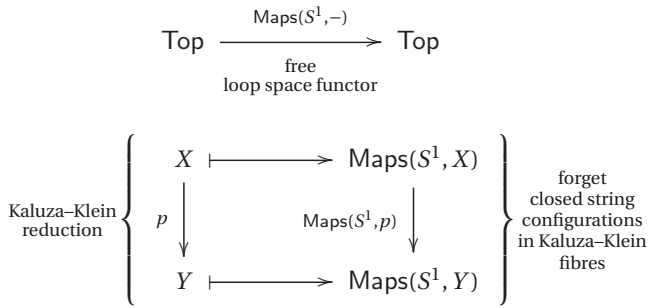

\begin{example}\label{ex:loop_space}
{\bf (Loop spaces and closed string configurations)}
In every category in which the collection of morphisms between any two objects itself carries structure that makes it the object of that category---for instance the topological space of maps $\sfMaps(\Sigma,X)$ between topological spaces $\Sigma$ and $X$---the assignment $X \mapsto \sfMaps(\Sigma,X)$ extends to a functor from the category to itself, see Figure~\ref{fig:FreeLoopSpaceFunctor}. If we think of $\Sigma = S^1$ as the spatial slice of a closed string world volume, and of a continuous function $X\to Y$ as a Kaluza--Klein compactification of space-time, then its image under this functor is the continuous function from the free loop space of $X$ to the free loop space of $Y$ which projects out from closed string configurations in $X$ their fibre components which get projected out under the Kaluza--Klein compactification.
\end{example}

\smallskip
Working with categories or relational sets instead of ordinary sets allows for a more detailed notion of distinction between objects. We can not only ask if two objects are equal, as in set theory, but we can also ask if two elements are related. It is only natural to want to extend this notion of distinction from objects to morphisms, which requires the introduction of relations between relations or morphisms between morphisms, which are also known as \emph{2-morphisms}. Iteration of this procedure leads then to the introduction of $n$-morphisms for any $n\in \mathbb{N}$. Such extended notions of categories are called \emph{higher categories} or {$n$-categories}, where $n$ labels the highest non-trivial level of $n$-morphism.

\begin{figure}[h]
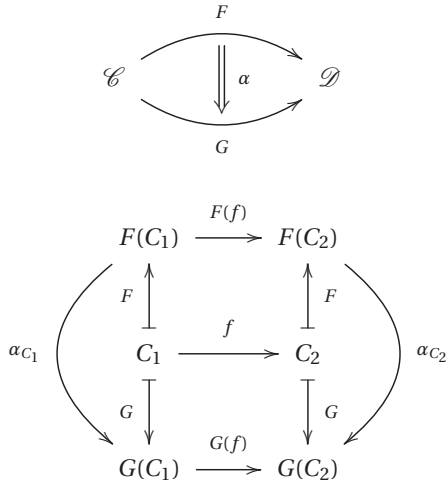

\begin{minipage}{8cm}
\vspace{-1cm}
\begin{eqnarray*}
&&\kern2.35cm\myxymatrix{
\scrC  \ar@/^4ex/[rr]^{F}="g1" \ar@/_4ex/[rr]_{G}="g2" \ar@{=>}^{\alpha} "g1"+<0ex,-3ex>;"g2"+<0ex,3ex> && \scrD
}\\
&&\kern1.1cm\myxymatrix{
F(C_1)  \ar[r]^{F(f)} \ar@/_8ex/[dd]_{\alpha_{C_1}} & F(C_2) \ar@/^8ex/[dd]^{\alpha_{C_2}}\\
C_1 \ar@{|->}[u]^{F}\ar@{|->}[d]_{G}\ar[r]^{f} & C_2\ar@{|->}[u]_{F}\ar@{|->}[d]^{G} \\
G(C_1) \ar[r]^{G(f)} & G(C_2)
}
\end{eqnarray*}
\end{minipage}
\vspace{-.7cm}
\caption{Definition of a natural transformation $\alpha:F\Rightarrow G$ between two functors $F,G:\scrC\to\scrD$ between two categories $\scrC$ and $\scrD$. The bottom diagram represents the component version of the top diagram, exhibiting the fact that $\alpha$ is encoded in a particular map $\mathsf{Obj}(\scrC)\rightarrow \mathsf{Mor}(\scrD)$.}
\label{fig:NaturalTransformation}
\vspace{-.7cm}
\end{figure}

While we shall return to higher categories below, it is important to notice that the category of small categories $\mathsf{Cat}$ is naturally a 2-category and the study of its 2-morphisms, called \emph{natural transformations}, was the original motivation for developing category theory, see also~\cite[page 1]{Freyd:1964baa}. Given two functors $F,G:\scrC\to\scrD$ between two categories $\scrC$ and $\scrD$, a natural transformation $\alpha:F\Rightarrow G$ transforms the functor $F$ into the functor $G$, respecting the internal structure as displayed in Figure~\ref{fig:NaturalTransformation}. Thus, $\mathsf{Cat}$ becomes a 2-category.

\begin{example}\label{ex:evaluation_map}
{\bf (Evaluation map)}
 Let us return to Examples~\ref{ex:wrapped_branes} and~\ref{ex:loop_space} and consider the composition of these functors. Given a closed string configuration space $\sfMaps(S^1, X)$ as in Figure~\ref{fig:FreeLoopSpaceFunctor}, we may extract from the pair of a world sheet field $\phi$ and a point $\sigma \in S^1$ on the circle the value of that world sheet field at that point. This is the \emph{evaluation map},
\begin{equation}
\begin{aligned}
\eps_X\,:\,\sfMaps(S^1,X)\times S^1\ &\to\ X~,\\
(\phi,\sigma)\ &\mapsto\ \phi(\sigma)~.
\end{aligned}
\end{equation}
As the space-time $X$ varies, these evaluation maps can be understood as the components of a natural transformation, see Figure~\ref{fig:NaturalTransformation}, from the composite of the two functors of Figures~\ref{fig:WrappedBraneFunctor} and~\ref{fig:FreeLoopSpaceFunctor} to the identity functor on topological spaces:
\begin{equation}\label{eq:NaturalEvaluation}
\kern-6pt\myxymatrix{
\sfTop  \ar@/^4ex/[rr]^{\sfMaps(S^1,-)\times S^1}="g1" \ar@/_4ex/[rr]_{\id}="g2" \ar@{=>}^{\eps} "g1"+<0ex,-3ex>;"g2"+<0ex,3ex> && \sfTop
}
\end{equation}
\end{example}

\smallskip
Let us make two more remarks concerning natural transformations. Firstly, a natural transformation $\alpha$ which has a left- and right-inverse or, equivalently, for which the component maps $\alpha_{(-)}:\mathsf{Obj}(\scrC)\rightarrow \mathsf{Mor}(\scrD)$ are isomorphisms for each object is called a \emph{natural isomorphism}. Secondly, any two given categories $\scrC$ and $\scrD$, the functors between them and the natural transformations between those form the \emph{functor category} $\mathsf{Fun}(\scrC,\scrD)$.

Natural transformations play a key role in our discussion since they allow us to define \emph{categorical equivalence}, \emph{adjoint pairs}, and \emph{adjoint equivalences} which correspond to the appropriate notions of `sameness' of categories, dualities, and duality equivalences, respectively. We briefly review the three in the following.

We often call two mathematical objects `essentially the same' or isomorphic, if we have an invertible map between them. Invertibility of a morphism amounts to the existence of another morphism so that left and right compositions of both equal the identity map. In the 2-category $\mathsf{Cat}$, it makes sense to loosen these equalities to \emph{relations} with the identity map, which yields the concept of `weak inverses'. We thus call two categories $\scrC$ and $\scrD$ \emph{equivalent}, if we have a pair of functors $F$ and $G$,
\begin{equation}
\kern-6pt\myxymatrix{
\mathcal{D}\ar@{->}@<-2pt>[rr]_-G\ar@{<-}@<+2pt>[rr]^-F&&\mathcal{C}
 }
\end{equation}
together with natural isomorphisms
\begin{equation}
 G\circ F\cong \id_\scrC~~~\mbox{and}~~~F\circ G\cong \id_\scrD~,
\end{equation}
rendering $F$ and $G$ weak inverses of each other.

Next, we turn to a generalisation of this picture. An \emph{adjunction} between two categories $\scrC$ and $\scrD$ is a pair of functors $L:\scrC\to\scrD$ and $R:\scrD\to\scrC$ equipped with natural transformations $\eta:\id_\scrC\Rightarrow R\circ L$ and $\eps: L\circ R\Rightarrow \id_\scrD$, called the \emph{adjunction unit} and \emph{adjunction counit}, respectively, such that so-called \emph{triangle identities}
\begin{equation}
 \begin{aligned}
&\kern-6pt\myxymatrix{
& (L\circ R\circ L)(-) \ar[dr]^{\eps_{L(-)}} &\\
L(-) \ar@{=}[rr]_{\id_{L(-)}}  \ar[ur]^{L(\eta_{(-)})}  && L(-)
}\\[5pt]
&\kern-6pt\myxymatrix{
R(-) \ar[dr]_{\eta_{R(-)}} \ar@{=}[rr]^{\id_{R(-)}} && R(-)\\
& (R\circ L\circ R)(-) \ar[ur]_{R(\eps_{(-)})} &
}
 \end{aligned}
\end{equation}
hold. These identities are simply the usual counit-unit relations,
\begin{equation}
 \begin{aligned}
  \id_\scrC &= \eps_{L(-)}\circ L(\eta_{(-)})~,\\
  \id_\scrD &= R(\eps_{(-)})\circ \eta_{R(-)}
 \end{aligned}
\end{equation}
(which correspond to the `zigzag-identities' in the ambient 2-category). We say that $L$ is \emph{left-adjoint} to $R$, $R$ is \emph{right-adjoint} to $L$, and $L$ and $R$ form an adjoint pair denoted by $L\dashv R$. We also write
\begin{equation}\label{eq:AdjointFunctors}
\kern-6pt\myxymatrix{
\mathcal{D}\ar@{->}@<-8pt>[rr]_-R^-{\bot}\ar@{<-}@<+8pt>[rr]^-L&&\mathcal{C}
}.
\end{equation}

One of the basic theorems of category theory shows (see e.g.~\cite[Proposition 1.39]{Schreiber:2018lab}) that two functors $L \dashv R$ being adjoint to each other means equivalently that there is a natural identification of morphisms `from $L(-)$' with those `into $R(-)$'. For $C\in \mathsf{Obj}(\scrC)$ and $D\in\mathsf{Obj}(\scrD)$, we then have
\begin{equation}\label{eq:AdjunctionHomIsomorphism}
\hom_\scrD(L(C),D)\cong \hom_\scrC(C,R(D))~,
\end{equation}
which gives a manifest expression of how $L$ and $R$ are \emph{dual} to each other (and motivates the nomenclature).

\smallskip
\begin{example}
{\bf (Left-adjoint functor)}
As a simple example, consider the functor $L:\sfSet\to\sfGrp$ that assigns to each set the free group generated by the elements of the set and let $R:\sfGrp\to\sfSet$ be the (forgetful) functor that assigns to each group its underlying set. Then, the functor $L$ is left-adjoint to $R$.
\end{example}

\smallskip
In the special case when both the unit $\eta$ and counit $\eps$ are natural isomorphisms, we obtain an \emph{adjoint equivalence}. Clearly, an adjoint equivalence is a special case of an equivalence of categories. One can show that any categorical equivalence can be improved to an adjoint equivalence.

\begin{example}\label{ex:top_adjoint_pair} {\bf (Adjunction)}
Recall that in Example~\ref{ex:evaluation_map}, we constructed a natural transformation between the composition of the two functors
\begin{equation}
  \sfTop \xrightarrow{~\sfMaps(S^1,-)~} \sfTop~~~\mbox{and}~~~
  \sfTop \xrightarrow{~(-)\times S^1~} \sfTop~,
\end{equation}
which we called $\eps$, suggesting the role of a counit in an adjunction. There is also a natural transformation for the inverse composition of the functors, which is constructed as follows. Every space-time $X$ has a canonical map into the space of closed string configurations with the result of adding a compact dimension to $X$. Concretely, it is given by sending each $x \in X$ to the string configuration that wraps the circle fibre over $x$,
\begin{equation}
\begin{aligned}
\eta_X\,:\,X\ &\to\ \sfMaps(S^1,X\times S^1)~,\\
x\ &\mapsto\ (\sigma\mapsto(x,\sigma))~.
\end{aligned}
\end{equation}
As in Example~\ref{ex:evaluation_map}, varying the space-time $X$ yields the natural transformation
\begin{equation}\label{eq:NaturalWrapping}
\kern-6pt\myxymatrix{
\sfTop  \ar@/^4ex/[rr]^{\id}="g1" \ar@/_4ex/[rr]_{\sfMaps(S^1,(-)\times S^1)}="g2" \ar@{=>}^{\eta} "g1"+<0ex,-3ex>;"g2"+<0ex,3ex> && \sfTop
}
\end{equation}

Altogether, the pair of natural transformations that extract string embedding fields~\eqref{eq:NaturalEvaluation} and assign wrapping modes~\eqref{eq:NaturalWrapping} exhibits the pair of functors that assigns wrapped brane configurations, see Figure~\ref{fig:WrappedBraneFunctor}, and performs Kaluza--Klein compactifications, see Figure~\ref{fig:FreeLoopSpaceFunctor}, as being adjoint to each other (e.g. \cite[Proposition 3.41]{Schreiber:2016laa}):\footnote{More precisely, here we understand $\sfTop$ as the category of \emph{compactly generated} topological spaces. See e.g.~\cite[Definition 3.35]{Schreiber:2016laa} for details.}
\begin{equation}
\kern-6pt\myxymatrix{
\sfTop \ar@<-8pt>[rr]_{\sfMaps(S^1,-)}^{\bot} && \sfTop \ar@<-8pt>[ll]_{(-)\times S^1}
}
\end{equation}
\end{example}

This concludes our lightening review of basic category theory. Next, we shall apply them in the context of dualities.

\subsection{Duality: a categorical point of view}\label{sec:duality}

\begin{figure}[h]
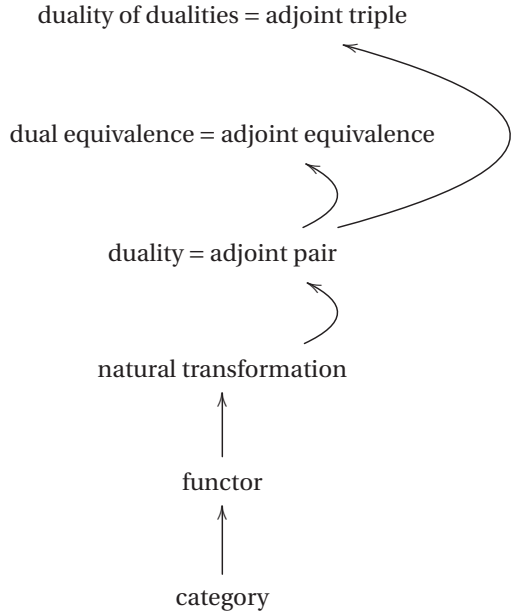

\vspace{-.8cm}
\begin{equation*}
\kern.7cm\myxymatrix{
\mbox{duality of dualities} = \mbox{adjoint triple}\\
\mbox{dual equivalence} = \mbox{adjoint equivalence}\\
\mbox{duality} = \mbox{adjoint pair} \ar@/_10ex/[u] \ar@/_25ex/[uu] \\
\mbox{natural transformation} \ar@/_10ex/[u]\\
\mbox{functor} \ar[u]\\
\mbox{category} \ar[u]
}
\end{equation*}
\vspace{-.7cm}
\caption{The hierarchy of concepts in category theory. Historically, the concepts of category theory were introduced by algebraic topologists in order to make sense of the \emph{natural transformations} which they saw in cohomology theory. Similarly, mathematically inclined string theorists could have invented category theory as the minimum mathematical context for speaking about \emph{duality}.}
\label{fig:Concepts}
\vspace{-.4cm}
\end{figure}

\emph{Duality} is certainly an ancient notion in philosophy which famously makes a curious reappearance in string theory. It was noticed some time ago \cite{Lambek:1981:111-121} that an excellent candidate to make precise the idea of duality is the mathematical concept of adjunctions, which, as seen above, are formulated in terms of the concept of natural transformation. Typically these are introduced as concepts in category theory but as the founding fathers of category theory noticed, the theory is really the minimum backdrop to speak about natural transformations \cite[page 1]{Freyd:1964baa}, hence about adjunction \cite[page 3]{Lawvere:1969aaa}, and hence about duality \cite{Lambek:1981:111-121} in the first place, see Figure~\ref{fig:Concepts}.

Some of the dualities in string theory are realised as equivalences of categories and as such often not advertised as adjunctions between categories. As mentioned above, however, these equivalences can be improved to duality or adjoint equivalences. Examples of string theoretic dualities that are expressed as equivalences of categories include mirror symmetry~\cite{Kontsevich:9411018}, see also~\cite{Aspinwall:2009isa}, and S-duality in the guise of geometric Langlands duality~\cite{Kapustin:2006pk}, see also~\cite{Frenkel:0906.2747}.

Let us start by discussing an important example from mathematics, namely the duality between spaces and their algebras of functions. Given any flavour of geometry (say differential geometry) and chosen coefficient space (say the real line) there is a functor which sends spaces to their algebras of functions and sends morphisms $f$ between spaces to the pre-composition with these maps,
\begin{equation} \label{eq:FunctorsApplyingAlgebrasOfFunctions}
\kern-6pt\myxymatrix{
\sfSpac \ar[rr]^{\substack{\text{functor that}\\[1pt] \text{assigns algebras}\\ \text{of functions}}} && \sfAlg^{\text{op}}\\
X \ar@{|->}[rr]^{(-)} \ar[d]_{f} &&\text{Functions}(X)\\
Y \ar@{|->}[rr]^{(-)} &&\text{Functions}(Y)\ar[u]_{(-)\circ f}
}
\end{equation}
Here, on the right-hand-side, we have the opposite category of algebras since a map between spaces induces the pullback of functions which goes in the opposite direction.

\begin{table*}
\centering
\begin{proptabular}[1cm]{c|c|c}{The duality between geometry and algebra is witnessed by fully faithfulness of functors assigning algebras of functions.}
\label{tab:IsbellDuality}
Geometry & $\myxymatrix{\text{Category}\ar[rr]^{\substack{\text{functor that}\\[1pt] \text{assigns algebras}\\ \text{of functions}}~~~~} &&\text{Dual Category}}$ & Algebra \\
topology & $\myxymatrix{\sfTop_{\text{compact Hausdorff}}\ar@{^{(}->}[rr]^{~~\text{Gelfand--Kolmogorov}}&&\sfAlg^{\text{op}}_{\text{commutative}}}$ & commutative algebra\\
topology & $\myxymatrix{\sfTop_{\text{compact Hausdorff}}\ar[rr]^{\text{Gelfand duality}}&&C^*\sfAlg^{\text{op}}_{\text{commutative}}}$ & commutative $C^*$-algebra\\
non-commutative topology & $\myxymatrix{\text{NC}\sfTop\ar@{=}[rr]^{:=}&&C^*\sfAlg^{\text{op}}}$ & non-commutative $C^*$-algebra\\
algebraic geometry & $\myxymatrix{\sfScheme_{\text{affine}}\ar@{=}[rr]^{:=}&&\sfAlg^{\text{op}}_{\text{finite commutative}}}$ & finitely generated commutative algebra\\
algebraic geometry & $\myxymatrix{\text{NC}\sfScheme_{\text{affine}}\ar@{=}[rr]^{~~~~~:=}&&\sfAlg^{\text{op}}_{\text{finite}}}$ & finitely generated non-commutative algebra\\
differential geometry & $\myxymatrix{\sfMfd\ar@{^{(}->}[rr]^{\text{Milnor's exercise}~~~~~~~}&&\sfAlg^{\text{op}}_{\text{commutative}}}$ & commutative algebra\\
super differential geometry & $\myxymatrix{\sfSMfd\ar@{^{(}->}[rr]&&\sfAlg^{\text{op}}_{\text{supercommutative}}}$ & supercommutative algebra\\
higher super geometry & $\myxymatrix{\mathsf{SL}_\infty\mathsf{Algebroid}\ar@{^{(}->}[rr]&&\text{dgc}\sfAlg^{\text{op}}_{\text{supercommutative}}}$ & differential graded supercommutative algebra\\
$\vdots$ & $\vdots$ & $\vdots$
\end{proptabular}
\end{table*}

\smallskip
\begin{example}
{\bf (Smooth manifolds)}
As a concrete example, consider the real algebra of smooth real-valued functions $\scrC^\infty(X,\IR)$ on some
smooth manifold $X$. The functor~\eqref{eq:FunctorsApplyingAlgebrasOfFunctions} reads explicitly as
\begin{equation} \label{eq:SmoothFunctionsFunctor}
\kern-6pt\myxymatrix{
\sfMfd \ar[rr]^{\scrC^\infty(-,\IR)} && \sfAlg^{\text{op}}_\IR\\
X \ar@{|->}[rr]^{\scrC^\infty(-,\IR)} \ar[d]_{f} &&\scrC^\infty(X,\IR)\\
Y \ar@{|->}[rr]^{\scrC^\infty(-,\IR)} &&\scrC^\infty(Y,\IR)\ar[u]_{\phi\,\mapsto\,\phi\circ f}
}
\end{equation}
This functor has a remarkable property which is of profound relevance for the mathematical formulation of physics: clearly only very special objects (algebras) are in its image but for those that are, it establishes a \emph{bijection} between the smooth functions between manifolds and algebra homomorphisms between their algebras of functions
\begin{equation}
\begin{aligned}
&\Big\{\,\sfMfd\ni X \xrightarrow{~~~~~f\in\scrC^\infty(X,Y)~~~~~} Y\in\sfMfd\,\Big\}\cong\\
&\kern.5cm\cong \Big\{\,\sfAlg^{\text{op}}_\IR\ni \scrC^\infty(X,\IR) \xleftarrow{~~~~~f^*~~~~~} \scrC^\infty(Y,\IR)\in\sfAlg^{\text{op}}_\IR\,\Big\}
\end{aligned}
\end{equation}
This statement is sometimes referred to as \emph{Milnor's exercise}, see \cite[Sections 35.8--35.10]{KolarMichorSlovak:1993baa}. By the structuralism established by category theory, which, recall, says that the nature of objects of a category is entirely reflected in the system of morphisms that relates it to other objects, this implies that for all practical purposes the algebras of functions $\scrC^\infty(X,\IR)$ with morphisms between them read in reverse are a complete stand-in for smooth manifolds $X$.
\end{example}

In general, a functor with the property that it induces bijections between the sets of morphisms
\begin{equation}\label{eq:FullEmbedding}
\kern-6pt\myxymatrix{
\scrC  \ar@{^{(}->}[rr]^-{F} && \scrD\\
\Big\{c_1\xrightarrow{~f~} c_2\Big\} \ar[rr]^-{F_{c_1,c_2}}_-{\cong} && \Big\{F(c_1)\xrightarrow{~F(f)~} F(c_2)\Big\}
}
\end{equation}
is called \emph{fully faithful} or a \emph{full embedding}. Again, by the structuralism this means that if we retain all the objects in the image of $F$ with all their morphisms to carve out a new category $\im(F)$ inside $\scrD$, then a fully faithful functor $F$ exhibits $\scrC$ and $\im(F)$ as being \emph{equivalent categories}, and we have
\begin{equation}
\kern-6pt\myxymatrix{
\scrC  \ar@{^{(}->}[rr]^-{F} \ar[dr]_{\cong} && \scrD\\
&\im(F) \ar@{^{(}->}[ur]&
}
\end{equation}

Applying this general fact of category theory to functors~\eqref{eq:FunctorsApplyingAlgebrasOfFunctions} that assign algebras of functions to spaces in given flavours of geometry means that as soon as these  functors happen to be fully faithful, they exhibit a \emph{dual equivalence} between geometric spaces and the corresponding kinds of algebras,
\begin{equation}
\kern-6pt\myxymatrix{
\sfSpac \ar@{^{(}->}[rr]^{\substack{\text{fully faithful functor }\\[1pt] \text{that assigns}\\[1pt] \text{algebras of functions}}} \ar[dr]_{\cong} && \sfAlg^{\text{op}}\\
&\sfGoodAlg^{\text{op}} \ar@{^{(}->}[ur]&
}
\end{equation}

A wealth of types of geometries are characterised via their algebras of functions by such fully faithful functors (see Table~\ref{tab:IsbellDuality}). This is what underlies the tremendous success of theoretical physics in discovering and handling new types of geometries---notably supergeometry, non-commutative geometry, and differential graded geometry, by deforming and handling algebras of functions.

\smallskip
\begin{example}\label{exa:DoubleDimRed}
{\bf (Double dimensional reduction)}\footnote{By `double dimensional reduction' we mean a simultaneous dimensional reduction of the world volume and the space-time.}
Consider now again the adjoint pair of Example~\ref{ex:top_adjoint_pair}. The corresponding adjunction/duality equivalence~\eqref{eq:AdjunctionHomIsomorphism} becomes
\begin{equation}\label{eq:ForMapsAdjunctionHomIsomorphism}
\underbrace{\Big\{X\times S^1\ \xrightarrow{~~}\ Y\Big\}}_{\substack{\text{maps from space-time}\\ \text{with an extra}\\[1.5pt] \text{compact direction}}}
\xleftrightarrow{~~~~~\cong~~~~~}\
\underbrace{\Big\{X\ \xrightarrow{~~}\ \sfMaps(S^1,Y)\Big\}}_{\substack{\text{maps into}\\ \text{free loop space}\\ \text{of coefficient space}}}~.
\end{equation}
In fact, this isomorphism applies not just to sets of maps, but also to sets of their \emph{homotopy classes}. We will discuss this in somewhat more detail in Section~\ref{sec:HomotopyTheory}.

Let us now specialise to the case where the coefficient space $Y$ is the Eilenberg--MacLane space \footnote{Recall that a connected topological space $X$ is said to be an \emph{Eilenberg--MacLane space of type $K(\sfG, n)$} whenever the $n$-th homotopy group of $X$ is isomorphic to $\sfG$ and all other homotopy groups are trivial. It can be shown that when $n > 1$, then $\sfG$ must be Abelian.} $K(\IZ,p+2)$, which is the classifying space for ordinary cohomology. This means that
\begin{equation}
\Big[ X \times S^1\ \xrightarrow{~~}\ K(\IZ,p+2)\Big]\cong H^{p+2}(X\times S^1,\IZ)
\end{equation}
is the ordinary cohomology of $X \times S^1$ that measures topological classes of flux fields to which $p$-branes may couple; here, square brackets on the left-hand-side denote homotopy equivalence classes. For instance, for $p =1$, this measures the $B$-field flux on $X \times S^1$ to which the string couples while for $p =2$, it measures the $C$-field flux to which the membrane couples.

The general adjunction~\eqref{eq:ForMapsAdjunctionHomIsomorphism} yields in this case
\begin{equation}
\begin{aligned}
&\underbrace{\Big[ X \times S^1\ \xrightarrow{~~}\ K(\IZ,p+2)\Big]}_{\substack{(p+2)\text{-form flux}\\ \text{on space-time with an}\\ \text{extra compact dimension}}}\cong\\
&\kern.5cm\cong\Big[X\ \xrightarrow{~~}\ \sfMaps(S^1,K(\IZ,p+2))\Big]\\
&\kern.5cm\cong\underbrace{\underbrace{\Big[X\ \xrightarrow{~~}\ K(\IZ,p+2)\Big]}_{\cong H^{p+2}(X,\IZ)}\times\underbrace{\Big[X\ \xrightarrow{~~}\ K(\IZ,p+1)\Big]}_{\cong H^{p+1}(X,\IZ)}}_{\substack{(p+2)\text{-form and }(p+1)\text{-form fluxes}\\ \text{on lower-dimensional space-time}}}
\end{aligned}
\end{equation}
and hence,
\begin{equation}\label{eq:IsomorphismDD}
\underbrace{H^{p+2}(X\times S^1,\IZ)}_{\text{total flux}}\cong\underbrace{H^{p+2}(X,\IZ)}_{\substack{\text{non-wrapped}\\ \text{flux}}}\oplus\underbrace{H^{p+1}(X,\IZ)}_{\substack{\text{wrapped}\\ \text{flux}}}~.
\end{equation}
This exhibits the \emph{double dimensional reduction} on flux fields \cite[Section 4.2]{Mathai:2003mu}. The isomorphism itself
also follows as a special case of the Gysin exact sequence, hence of the Serre spectral sequence, but the derivation via adjunction~\eqref{eq:ForMapsAdjunctionHomIsomorphism} makes manifest \cite[Remark 3.9]{Fiorenza:2016oki} that this is really the construction of double dimensional reduction for branes in string theory \cite{Duff:1987bx}.

In fact, double dimensional reduction as an adjunction applies more generally, also for Kaluza--Klein compactification on non-trivial circle bundles and in superspace, such as for T-duality of Ramond--Ramond fluxes \cite{Fiorenza:2016oki}, and also including gauge enhancement, such as for M/IIA-duality \cite{Braunack-Mayer:2018uyy}.
This is reviewed in more detail in the contribution 
\cite{contrib:fiorenza} to this volume. 
\end{example}

\subsection{Homotopy theory and the gauge principle}\label{sec:HomotopyTheory}

Let us now return to a point we already made when introducing natural transformations and expand on it. Traditionally, mathematics and physics have been founded on set theory whose concept of sets is that of \emph{bags of distinguishable points}. However, fundamental physics is governed by the \emph{gauge principle} which says that it is meaningless to ask if any two given \emph{things} such such as two field histories $x$ and $y$ are equal. Instead, the right question to ask is if there is a \emph{gauge transformation} between them:
\begin{equation}\label{eq:AGaugeTransformation}
\kern-6pt\myxymatrix{
x \ar[rr]^{h}_{\cong} && y
}
\end{equation}
In mathematics this is called a \emph{homotopy}.

The gauge principle applies to gauge transformations/homotopies themselves,
and thus leads to gauge-of-gauge transformations or homotopies of homotopies
\begin{equation}\label{eq:AHigherGaugeTransformation}
\kern-6pt\myxymatrix{
x \ar@/^2pc/[rr]^{h_1}_{\ }="s" \ar@/_2pc/[rr]_{h_2}^{\ }="t" && y \ar@{=>}^{\chi} "s"; "t"
}
\end{equation}
and so on to ever higher gauge transformations or higher homotopies:
\begin{equation}\label{eq:HigherHigher}
\kern-6pt\myxymatrix{
x \ar@/^2pc/[rr]^{h_1}_{\ }="s" \ar@/_2pc/[rr]_{h_2}^{\ }="t" && y \ar@/^1pc/@{=>}^{\chi_1} "s"; "t" \ar@/_1pc/@{=>}_{\chi_2} "s"; "t" \ar@{=>} (12,0); (16,0) \ar@{-} (12,0); (16,0)
}
\end{equation}
A collection of elements with higher gauge transformations/higher homotopies is called a \emph{higher homotopy type}. Hence the theory of homotopy types, \emph{homotopy theory} \cite{Spalinski:1995:73-126}, see also\cite{Schreiber:2016laa}, is a mathematical foundation like set theory \cite{UFP2013:aa}, see also \cite{Shulman:1703.03007} but with the concept of gauge transformation built into it \cite{Shulman:2014:109-126}: \emph{homotopy theory is gauged mathematics.}

\begin{figure}[h]
\begin{center}
\tikzset{->-/.style={decoration={markings,mark=at position #1 with {\arrow{>}}},postaction={decorate}}}
\begin{tikzpicture}[scale=0.7,every node/.style={scale=0.7}]
\begin{scope}[xshift = 0cm, yshift= 20pt]
\draw[line width=.7pt] (0,2) -- (2,2);  \draw (-.4,2.4) node {$(0,1)$}; \draw (2.4,2.4) node {$(1,1)$};
\draw[line width=.7pt] (0,1.5) -- (2,1.5);
\draw[line width=.7pt] (0,1) -- (2,1);
\draw[line width=.7pt] (0,0.5) -- (2,0.5);
\draw[line width=.7pt] (0,0) -- (2,0); \draw (-.4,-.4) node {$(0,0)$}; \draw (2.4,-.4) node {$(1,0)$};
\draw[dashed,->,line width=.7pt] (2,0) -- (2,2);
\draw[dashed,->,line width=.7pt] (1.5,0) -- (1.5,2);
\draw[dashed,->,line width=.7pt] (1,0) -- (1,2);
\draw[dashed,->,line width=.7pt] (0.5,0) -- (0.5,2);
\draw[dashed,->,line width=.7pt] (0,0) -- (0,2);
\end{scope}
\draw [dashed,->,line width=.7pt] plot [smooth] coordinates {(7.25,.86) (7.15,1.5) (7.25,2.14) };
\draw [dashed,->,line width=.7pt] plot [smooth] coordinates {(5.75,.86) (5.9,1.5) (5.75,2.14) };
\draw [dashed,->,line width=.7pt] plot [smooth] coordinates {(6.5,0.5) (6.5,2.5) };
\draw [line width=.7pt] plot [smooth] coordinates {(5,1.5) (6.5,2.5) (8,1.5) };
\draw [line width=.7pt] plot [smooth] coordinates {(5,1.5) (6.5,2) (8,1.5) };
\draw [line width=.7pt] plot [smooth] coordinates {(5,1.5) (6.5,1.5) (8,1.5) };
\draw [line width=.7pt] plot [smooth] coordinates {(5,1.5) (6.5,1) (8,1.5) };
\draw [line width=.7pt] plot [smooth] coordinates {(5,1.5) (6.5,0.5) (8,1.5) };
\draw [] plot [smooth cycle] coordinates {(4,-1) (4.5,2) (4,4) (6,3.5) (9,4) (8.5,2) (10,-1) (7,-.5) };
\draw (6.5,3) node {$g$}; \draw (6.5,0) node {$f$};
\end{tikzpicture}
\vspace{-.5cm}
\caption{Homotopy in $\sfTop$ is given by the evident 1-parameter continuous deformation of continuous functions.}
\label{fig:HomotopyInTopologialSpaces}
\end{center}
\vspace{-.7cm}
\end{figure}
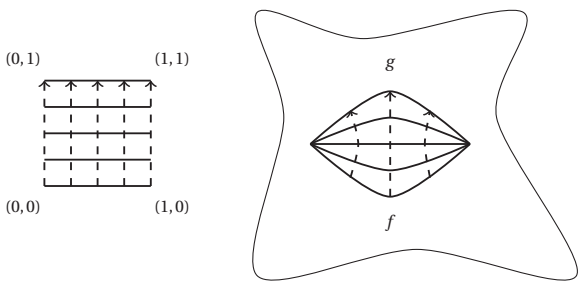

\begin{figure}[h]
\vspace{-.2cm}
\begin{eqnarray*}
&&\kern1.8cm\myxymatrix{
X\times\{0\}\cong X \ar[dr]^f \ar[d]_{\id_X\times\iota_0} & \\
X\times I\ar[r]^h & Y\\
X\times\{1\}\cong X \ar[ur]_g\ar[u]^{\id_X\times\iota_1} &
}\\
&&\kern1.6cm\myxymatrix{
&  \sfMaps(\{0\},Y)\cong Y \\
X \ar[ur]^f\ar[dr]_g\ar[r]^h & \sfMaps(I,Y) \ar[u]_{\sfMaps(\iota_0,Y)} \ar[d]^{\sfMaps(\iota_1,Y)}\\
& \sfMaps(\{1\},Y)\cong Y
}
\end{eqnarray*}
\vspace{-.7cm}
\caption{Homotopies/gauge transformations in general model categories are defined, by direct analogy with the topological case (Figure~\ref{fig:HomotopyInTopologialSpaces}) in terms of \emph{cylinder objects} $X \times I$ for $X$ or \emph{path space objects} $\sfMaps(I,Y)$ for $Y$, which may (but need not) come from an \emph{interval object}~\eqref{eq:IntervalObject} via the Cartesian product (Figure~\ref{fig:WrappedBraneFunctor}) and mapping spaces (Figure~\ref{fig:FreeLoopSpaceFunctor}). The upper picture defines a \emph{left homotopy} and the lower picture a \emph{right homotopy}, respectively.}
\label{fig:LeftRightHomotopies}
\vspace{-.4cm}
\end{figure}

Let $I:=[0,1]$ be the standard topological interval
with its standard endpoint inclusions
\begin{equation}\label{eq:IntervalObject}
\myxymatrix{
\ast \ar@{^{(}->}[d]_{\iota_0} \ar@{=}[dr] & \\
I \ar[r] & \ast\\
\ast \ar@{_{(}->}[u]^{\iota_1}\ar@{=}[ur]
}
\end{equation}
The standard realisation of homotopy in $\sfTop$ shown in Figure~\ref{fig:HomotopyInTopologialSpaces} is defined as in Figure~\ref{fig:LeftRightHomotopies} via \emph{cylinder spaces} $X \times I$ (as in Figure~\ref{fig:WrappedBraneFunctor})
and \emph{path spaces} $\sfMaps(I, X)$ (as in Figure~\ref{fig:FreeLoopSpaceFunctor}).

In order to speak of homotopies or gauge transformations in more generality than this, one simply generalises this concept of cylinders and path spaces: a category $\scrC$ with a good supply of \emph{cylinder objects} $X \times I$ and path space objects $\sfMaps(I,X)$
inducing left/right homotopies as in Figure~\ref{fig:LeftRightHomotopies} is called a \emph{model category} which is short for \emph{category of models for homotopy types} \cite{Quillen:1967bcaa}, see e.g.~\cite[Definition 3.3]{Spalinski:1995:73-126} and \cite[Definition 2.3]{Schreiber:2016laa}.

Every model category $\scrC$ comes with its \emph{homotopy category} (or \emph{derived category}),
\begin{equation}\label{eq:HomotopyCategory}
\mathsf{Ho}(\scrC):=\left\{~\substack{\text{the good objects of }\scrC\\ \text{and the homotopy classes of morphisms}}~\right\},
\end{equation}
whose objects are good models for homotopy types and whose morphisms are homotopy classes of morphisms in $\scrC$.

\begin{example}
{\bf (Topological homotopy theory)}
The ar\-chetypical example of homotopy theory, often understood by default, is that presented by topological spaces: the category $\sfTop$ of topological spaces,c which we already encountered in our list of concrete categories in Table~\ref{tab:ConcreteCategories}, becomes the model category $\sfTop_{\text{Quillen}}$ via the usual cylinder spaces and path spaces as in Figure~\ref{fig:HomotopyInTopologialSpaces}. This is the foundational result of Quillen's~\cite{Quillen:1967bcaa},  see also~\cite[Section 1]{Schreiber:2016laa}. Beware, however, that when regarded in topological homotopy theory, topological spaces lose their geometric nature (their topology) which now just serves to conveniently \emph{present} a homotopy type, i.e.~a system of higher order gauge transformations. The combination of actual geometry (topological, differential geometric
or super geometric) with homotopy theory is the topic of \emph{higher geometry} discussed in Section~\ref{sec:HigherGeometry} below.
\end{example}

\begin{example}
{\bf (Chain homotopy theory)}
Various \emph{shadows} of higher homotopy theory are more widely familiar. An important example is the homotopy theory of \emph{chain complexes} also known as \emph{homological algebra}, see e.g.~\cite{Weibel:1994aa} and in particular~\cite{Schreiber:2012lnaa} for this point of view.

Consider a chain complex $V_\bullet$ in non-negative degree,
\begin{equation}
\cdots\ \xrightarrow{~\partial~}\ V_2\ \xrightarrow{~\partial~}\ V_1\ \xrightarrow{~\partial~}\ V_0
\quad\mbox{with}\quad
\partial\circ\partial=0~.
\end{equation}
Let $x,y \in V_0$ be two elements in degree 0 whose images in degree-0 chain homology $[x], [y]\in H_0(V_\bullet):=V_0/\im(\partial|_{V_1})$ are equal, $[x]=[y]$. This means that there exists an element $h \in V_1$ in degree 1 such that the difference $y-x$ is a coboundary $\partial h = y - x$. Hence, every such $h$ is a reason for the equality $[x]=[y]$ and so, a homotopy $\big(\partial h = y - x\big)~\longleftrightarrow~\big(x\ \xrightarrow{~h~}\ y\big)$. Next, let $h_{1,2} \in V_1$ be two coboundaries between $x$ and $y$. Then, an element $\chi\in V_2$ with $\partial \chi = h_2 - h_1$ is a way for them to be equal by means of a homotopy between homotopies~\eqref{eq:AHigherGaugeTransformation}. Specifically, this is a way for them to be equal in degree 1 chain homology $H_1(V_\bullet):=\ker(\partial|_{V_1})/\im(\partial|_{V_2})$ in the sense that $[0] = [h_2-h_1]$ witnessed by $0\ \xrightarrow{~\chi~}\ (h_2-h_1)$. It is now evident how this generalises to third-order homotopy as in~\eqref{eq:HigherHigher} or even to higher order.

In this way, every chain complex defines a homotopy type. This is called the \emph{Dold--Kan correspondence}, see e.g.~\cite[Section III.2]{Goerss:1999aa}.
\end{example}

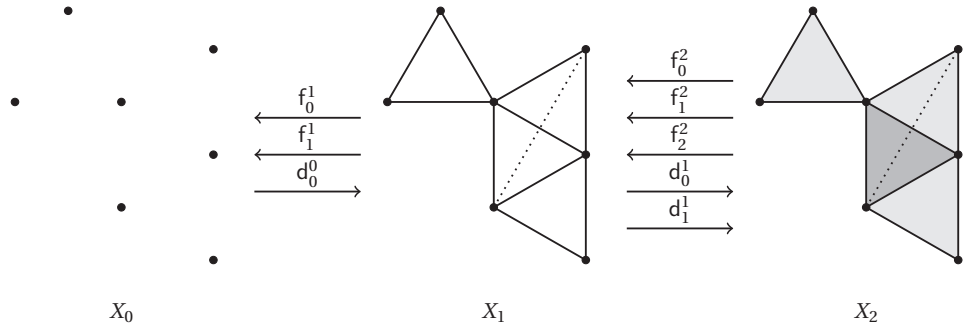
\begin{figure*}[h]
\begin{center}
\tikzset{->-/.style={decoration={markings,mark=at position #1 with {\arrow{>}}},postaction={decorate}}}
\begin{tikzpicture}[scale=0.7,every node/.style={scale=.9}]
\filldraw [] (3.73,-3) circle (2pt);
\filldraw [] (2,-2) circle (2pt);
\filldraw [] (3.73,-1) circle (2pt);
\filldraw [] (3.73,1) circle (2pt);
\filldraw [] (1,1.73) circle (2pt);
\filldraw [] (2,0) circle (2pt);
\filldraw [] (0,0) circle (2pt);
\filldraw [] (10.73,-3) circle (2pt);
\filldraw [] (9,-2) circle (2pt);
\filldraw [] (10.73,-1) circle (2pt);
\filldraw [] (10.73,1) circle (2pt);
\filldraw [] (8,1.73) circle (2pt);
\filldraw [] (9,0) circle (2pt);
\filldraw [] (7,0) circle (2pt);
\draw[dotted,line width=.7pt] (9,-2) -- (10.73,1);
\draw[line width=.7pt] (10.73,1) -- (10.73,-1);
\draw[line width=.7pt] (10.73,-3) -- (10.73,-1);
\draw[line width=.7pt] (9,-2) -- (10.73,-3);
\draw[line width=.7pt] (9,-2) -- (10.73,-1);
\draw[line width=.7pt] (9,0) -- (10.73,-1);
\draw[line width=.7pt] (9,0) -- (9,-2);
\draw[line width=.7pt] (9,0) -- (10.73,1);
\draw[line width=.7pt] (8,1.73) -- (9,0);
\draw[line width=.7pt] (7,0) -- (8,1.73);
\draw[line width=.7pt] (7,0) -- (9,0);
\filldraw[color=gray!20] (14,0) -- (16,0) -- (15,1.73) -- cycle;
\filldraw[color=gray!20] (16,0) -- (17.73,-1) -- (17.73,1) -- cycle;
\filldraw[color=gray!60] (16,0) -- (16,-2) -- (17.73,-1) -- cycle;
\filldraw[color=gray!20] (16,-2) -- (17.73,-3) -- (17.73,-1) -- cycle;
\filldraw [] (17.73,-3) circle (2pt);
\filldraw [] (16,-2) circle (2pt);
\filldraw [] (17.73,-1) circle (2pt);
\filldraw [] (17.73,1) circle (2pt);
\filldraw [] (15,1.73) circle (2pt);
\filldraw [] (16,0) circle (2pt);
\filldraw [] (14,0) circle (2pt);
\draw[dotted,line width=.7pt] (16,-2) -- (17.73,1);
\draw[line width=.7pt] (17.73,1) -- (17.73,-1);
\draw[line width=.7pt] (17.73,-3) -- (17.73,-1);
\draw[line width=.7pt] (16,-2) -- (17.73,-3);
\draw[line width=.7pt] (16,-2) -- (17.73,-1);
\draw[line width=.7pt] (16,0) -- (17.73,-1);
\draw[line width=.7pt] (16,0) -- (16,-2);
\draw[line width=.7pt] (16,0) -- (17.73,1);
\draw[line width=.7pt] (15,1.73) -- (16,0);
\draw[line width=.7pt] (14,0) -- (15,1.73);
\draw[line width=.7pt] (14,0) -- (16,0);
\draw[->,line width=.7pt] (6.5,-.3) -- (4.5,-.3); \draw (5.5,.05) node {$\sff^1_0$};
\draw[->,line width=.7pt] (6.5,-1) -- (4.5,-1); \draw (5.5,-.65) node {$\sff^1_1$};
\draw[->,line width=.7pt] (4.5,-1.7) -- (6.5,-1.7); \draw (5.5,-1.35) node {$\sfd^0_0$};
\draw[->,line width=.7pt] (13.5,.4) -- (11.5,.4); \draw (12.5,.75) node {$\sff^2_0$};
\draw[->,line width=.7pt] (13.5,-.3) -- (11.5,-.3); \draw (12.5,.05) node {$\sff^2_1$};
\draw[->,line width=.7pt] (13.5,-1) -- (11.5,-1); \draw (12.5,.-.65) node {$\sff^2_2$};
\draw[->,line width=.7pt] (11.5,-1.7) -- (13.5,-1.7); \draw (12.5,-1.35) node {$\sfd^1_0$};
\draw[->,line width=.7pt] (11.5,-2.4) -- (13.5,-2.4);  \draw (12.5,-2.05) node {$\sfd^1_1$};
\draw (2,-4) node {$X_0$}; \draw (9,-4) node {$X_1$}; \draw (16,-4) node {$X_2$};
\end{tikzpicture}
\end{center}
\vspace{-.7cm}
\caption{A simplicial set $X$ is a sequence of sets $X_n$ whose elements behave like the $n$-simplices inside $X$ with maps between them that behave like assigning faces of simplices  $\sff^n_i$ as well as assigning degenerate simplices $\sfd^n_i$. In homotopy theory simplicial sets provide a particularly useful tool for handling systems of higher order gauge transformations, as in~\eqref{eq:AGaugeTransformation}, \eqref{eq:AHigherGaugeTransformation}, and~\eqref{eq:HigherHigher}.}
\label{fig:IdeaOfSimplicialSets}
\end{figure*}

\begin{example}
\noindent {\bf (Poincar{\'e} lemma)} For $X$ a smooth $n$-dimensional manifold, its de Rham complex $(\Omega^\bullet(X),\dd)$ is the chain complex
\begin{equation}
\scrC^\infty(X)=\Omega^0(X)\ \xrightarrow{~\dd~}\ \Omega^1(X)\ \xrightarrow{~\dd~}\ \cdots\ \xrightarrow{~\dd~}\ \Omega^n(X)
\end{equation}
of differential forms with the de Rham differential $\dd$ between them. If we have two smooth functions $f,g:X\to Y$ between two smooth manifolds $X$ and $Y$ and a smooth homotopy $h:X\times[0,1]\to Y$ with $h(-,0)=f(-)$ and $h(-,1)=g(-)$ then there is a chain homotopy $h^*:f^*\Rightarrow g^*$ between the induced morphisms $f^*,g^*:(\Omega^\bullet(Y),\dd_Y)\to(\Omega^\bullet(X),\dd_X)$ and the corresponding de Rham cohomology groups coincide $H^\bullet_{\rm dR}(f^*)\cong H^\bullet_{\rm dR}(g^*)$.

In particular, when $X=Y=D^n$ with $D^n$ the open ball in $\IR^n$, then $0:D^n\to D^n$ and $\id:D^n\to D^n$ are homotopic by means of $h(x,t)=tx$ for $x\in D^n$ and $t\in[0,1]$. We may then conclude from the above that $H^0_{\rm dR}(D^n)\cong\IR$ and $H^p_{\rm dR}(D^n)=0$ for $p\geq1$.  This is, of course, the statement of the \emph{Poincar\'e lemma}. We may rephrase this as follows. Let $0\in D^n$. Then, the de Rham complex for $\{0\}$ is
\begin{equation}
\underbrace{\IR\ \xrightarrow{~0~}\ 0\ \xrightarrow{~0~}\ \cdots\ \xrightarrow{~0~}\ 0}_{n+1\text{ terms}}~.
\end{equation}
Furthermore, we trivially have the chain homotopy
\begin{subequations}\label{eq:PoincareChainEquivalence}
\begin{equation}
\kern-6pt\myxymatrix{
(\Omega^\bullet(\{0\}),0) \ar@{^{(}->}[r]\ar@/_9ex/[rr]_{\id} & (\Omega^\bullet(D^n),\dd)\ar@{=}[d] \ar@{->>}[r] & (\Omega^\bullet(\{0\}),0) \\
&
}
\end{equation}
The statement of the Poincar\'e lemma is then that we also have a chain homotopy the `other way around', that is,
\begin{equation}
\kern-6pt\myxymatrix{
(\Omega^\bullet(D^n),\dd) \ar@{->>}[r]\ar@/_9ex/[rr]_{\id} & (\Omega^\bullet(\{0\}),0)\ar@{=>}[d] \ar@{^{(}->}[r] & (\Omega^\bullet(D^n),\dd) \\
&
}
\end{equation}
\end{subequations}
Together, the diagrams~\eqref{eq:PoincareChainEquivalence} define a \emph{chain homotopy equivalence} between the de Rham complexes $(\Omega^\bullet(\{0\}),0)$ and $(\Omega^\bullet(D^n),\dd)$, respectively.
\end{example}

\smallskip
\begin{example}
{\bf (Differential-graded commutative algebraic super homotopy theory)} Due to its relevance to supergravities and their higher form fluxes, let us mention the following example, albeit with little explanation. The category of real differential graded commutative superalgebras carries a \emph{projective} model category structure $\mathsf{dgcSAlg}_{\mathrm{proj}}$ with path space objects given by tensor product of algebras with $\Omega^\bullet_{\mathrm{poly}}([0,1])$, see \cite{Bousfield:1976:0-0} and \cite[Theorems 1 \& 6.5]{Carchedi:1211.6134}. As pointed out in references \cite{Fiorenza:2013nha,Fiorenza:2016ypo,Fiorenza:2016oki,Braunack-Mayer:2018uyy,Huerta:2018xyh}, this provides the homotopy theory of what in the supergravity literature are called \emph{FDA}s or \emph{CIS}s: local models for higher dimensional supergravity on superspace \cite{vanNieuwenhuizen:1982zf,D'Auria:1982nx}, see also \cite{Castellani:1991et}.
\end{example}

\begin{example}\label{exa:SimplicialHomotopy}
{\bf (Simplicial homotopy theory)}
A precise and useful handle on general systems of higher order gauge transformations---as in~\eqref{eq:AGaugeTransformation},
\eqref{eq:AHigherGaugeTransformation}, and~\eqref{eq:HigherHigher}---is provided by \emph{simplicial homotopy theory}~\cite{Quillen:1967bcaa}, see also the text book~\cite{Goerss:1999aa}.\footnote{For a discussion within the context of higher principal bundles and higher gauge theory, see also~\cite{Jurco:2016qwv}.}

Here, a \emph{simplicial $n$-simplex} is an $n$-dimensional generalisation of a triangle; a \emph{simplicial complex} is what is obtained from glueing simplices of any dimension, and, generally a \emph{simplicial set} $X$ is defined by:
\smallskip
\begin{enumerate}[label=\roman*),leftmargin=*]
\item for each $n\in\IN\cup\{0\}$
a set $X_n$ of would-be \emph{$n$-simplices};
\item
with maps between these sets that behave like
\begin{enumerate}[label=\alph*),leftmargin=*]
\item sending $(n+1)$-simplices to their $n$-dimensional faces---the \emph{face maps} $\sff^n_i:X_n\to X_{n-1}$ for $i=0,\ldots,n$;
\item constructing degenerate $(n+1)$-simplices on given $n$-simplices---the \emph{degeneracy maps} $\sfd^n_i:X_n\to X_{n+1}$ for $i=0,\ldots,n$;
\end{enumerate}
\noindent and obey the \emph{simplicial identities}:
\begin{equation}
 \begin{aligned}
  \sff_i\circ\sff_j&=\sff_{j-1}\circ\sff_i\quad\mbox{for}\quad i< j~,\\
  \sfd_i\circ\sfd_j&=\sfd_{j+1}\circ\sfd_i\quad\mbox{for}\quad i\leq j~,\\
  \sff_i\circ\sfd_j&=\sfd_{j-1}\circ\sff_{i}\quad\mbox{for}\quad i< j~,\\
  \sff_i\circ\sfd_j&=\sfd_j\circ\sff_{i-1}\quad\mbox{for}\quad i>j+1~,\\
  \sff_i\circ\sfd_i&=\id=\sff_{i+1}\circ\sfd_i~,
 \end{aligned}
\end{equation}
\end{enumerate}
see Figure~\ref{fig:IdeaOfSimplicialSets}.

\smallskip

We may also view a simplicial set as a functor. Indeed, let $\Delta$ be the simplex category which is the category that has the finite totally ordered sets $[n]:=\{0,\dots,n\}$ for $n\in\IN\cup\{0\}$ as objects and order-preserving maps as morphisms. A simplicial set $X$ is then simply a functor $X:\Delta^{\rm op}\to\sfSet$, that is, a $\sfSet$-valued \emph{presheaf} on $\Delta$. The face and degeneracy maps follow the generators that generate the morphisms of $\Delta$.

The prime example of a simplicial set is the standard simplicial $n$-simplex $\Delta^n$ which is the functor $\sfhom_\Delta(-,[n]):\Delta^{\rm op}\to\sfSet$. Furthermore, the standard simplicial 1-simp\-lex $I:=\Delta^1$ serves as the interval object~\eqref{eq:IntervalObject} in simplicial homotopy theory (see Figure~\ref{fig:LeftRightHomotopies}). This corresponds to a model category of simplicial sets denoted by $\mathsf{sSet}_{\text{Quillen}}$. A slightly less trivial example of a simplicial set is the resulting \emph{cylinder object} $I \times \Delta^2$ on the 2-simplex shown in Figure~\ref{fig:SimplicialCylinderObject}.

\begin{figure}[h]
\begin{center}
\tikzset{->-/.style={decoration={markings,mark=at position #1 with {\arrow{>}}},postaction={decorate}}}
\begin{tikzpicture}[scale=0.7,every node/.style={scale=1}]
\filldraw[pattern=north west lines,pattern color=gray!30,draw=white] (0,5) -- (5,1) -- (3,4) -- cycle;
\filldraw[color=gray!40] (0,0) -- (5,1) -- (3,4) -- cycle;
\filldraw [] (5,6) circle (2pt);
\filldraw [] (3,4) circle (2pt);
\filldraw [] (0,5) circle (2pt);
\filldraw [] (5,1) circle (2pt);
\filldraw [] (3,-1) circle (2pt);
\filldraw [] (0,0) circle (2pt);
\draw[line width=.7pt] (3,4) -- (5,6);
\draw[line width=.7pt] (5,1) -- (5,6);
\draw[line width=.7pt] (3,-1) -- (5,1);
\draw[line width=.7pt] (3,-1) -- (3,4);
\draw[dotted,line width=.7pt] (0,5) -- (5,1);
\draw[line width=.7pt] (0,5) -- (5,6);
\draw[line width=.7pt] (0,5) -- (3,4);
\draw[line width=.7pt] (0,0) -- (5,1);
\draw[line width=.7pt] (0,0) -- (3,-1);
\draw[line width=.7pt] (0,0) -- (0,5);
\draw (-.3,5.05) node {$1'$}; \draw (3,4.4) node {$2'$}; \draw (5.3,6.1) node {$0'$};
\draw (-.3,0) node {$1$}; \draw (3,-1.3) node {$2$}; \draw (5.3,1) node {$0$};
\draw (2.5,0) node {$\Delta^2$}; \draw (-.8,2.5) node {$I=\Delta^1$};
\end{tikzpicture}
\end{center}
\vspace{-.7cm}
\caption{The canonical cylinder object on the simplicial 2-simplex in simplicial homotopy theory. Morphisms of simplicial sets out of this exhibit a left homotopy (as in Figure~\ref{fig:LeftRightHomotopies}) between its restrictions to the top and to the bottom simplicial 2-simplices (filled triangles).}
\label{fig:SimplicialCylinderObject}
\vspace{-.5cm}
\end{figure}
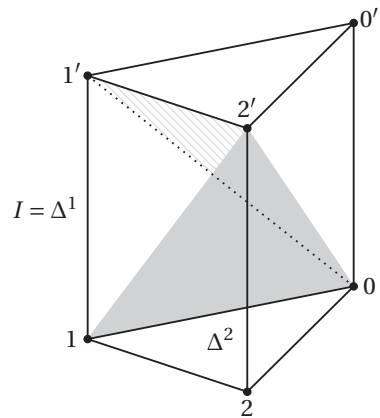

It is important to stress, however, that this has nothing to do with discretised space-time: whilst simplicial sets are indeed combinatorial concepts, they capture systems of higher gauge transformations as in~\eqref{eq:AGaugeTransformation}, \eqref{eq:AHigherGaugeTransformation}, and~\eqref{eq:HigherHigher}. It is certainly true that we may associate to every topological space $X$ its \emph{singular simplicial set}
\begin{equation}\label{eq:SingularSimplicialComplex}
\mathrm{Sing}(X)_n:=\ \Big\{~|\Delta^n|\ \xrightarrow{~~\text{continuous}~~}\ X~\Big\}
\end{equation}
and that this construction establishes an equivalence between topological and simplicial homotopy theory. It retains, however, no geometric information about $X$, and it is not meant to. Instead, the geometry of $X$ is retained and combined with homotopy theory/higher gauge theory in the theory of \emph{higher geometry} to which we turn in Section~\ref{sec:HigherGeometry}.
\end{example}

\smallskip
\begin{example}\label{ex:nerve} {\bf (Nerve of a category)} Each category $\scrC$ gives rise to a simplicial set, called its \emph{nerve}. The 0-simplices are the objects, the simplicial 1-simplices the morphisms and the simplicial $n$-simplices are $n$-tuples of composable morphisms. The face maps $\sff^n_{0,1}$ are projections onto subtuples for $n>1$ and onto source/domain and target/image of the morphisms for $n=1$. The remaining face maps $\sff^n_i$ correspond to composing morphisms $i$ and $i+1$ in the tuple. The degeneracy maps $\sfd^n_i$ are the obvious insertions of the identity morphisms in the category.
\end{example}

\smallskip
A simplicial set as in Figure~\ref{fig:IdeaOfSimplicialSets} is called a \emph{Kan complex} or \emph{$\infty$-groupoid} if its simplicial $n$-simplices may be composed and each has an inverse under composition both up to simplicial $(n+1)$-simplices. It turns out that $\infty$-groupoids are the \emph{good models} for simplicial homotopy types.

\smallskip
\begin{example}\label{ex:delooping} {\bf (Delooping groupoid)} For $\sfG$ a discrete group its \emph{delooping groupoid} is the simplicial set $\mathbf{B}\sfG$ (which happens to be an $\infty$-groupoid) which is defined by:
\smallskip
\begin{enumerate}[label=\roman*),leftmargin=*]
\item a single simplicial 0-simplex $\ast$;
\item simplicial 1-simplices $g:\ast\to\ast$ corresponding to the group elements $g\in\sfG$;
\item simplicial 2-simplices corresponding to the products in the group, that is,
\vspace{-10pt}
\begin{equation}
\begin{aligned}
\xymatrix{
&\ast\ar[dr]^{g_2}&\\
\ast\ar[ur]^{g_1}\ar[rr]_{g_2\cdot g_1} && \ast
}
\end{aligned}
\end{equation}
\vspace{-15pt}
\item simplicial 3-simplices witnessing associativity of the product;
\item and so on.
\end{enumerate}
Note that this simplicial set is the nerve of a category with one object $\ast$ and a set of morphisms $\sfG$.
\end{example}

\smallskip
An adjunction between model categories is called a \emph{Quillen adjunction}
\begin{equation}
\kern-6pt\myxymatrix{
\scrD \ar@<-10pt>[rr]_{R}^{\bot_{\text{Quillen}}} && \scrC \ar@<-10pt>[ll]_{L}
}
\end{equation}
if, very roughly (see \cite[Definition 2.46, Lemma 2.48]{Schreiber:2016laa} for details) the left adjoint functor $L$ preserves cylinder objects and the right adjoint functor $R$ preserves path space objects (as in Figure~\ref{fig:LeftRightHomotopies}). This, in turn, induces (see \cite[Proposition 2.49]{Schreiber:2016laa}) an adjunction of derived functors on homotopy categories~\eqref{eq:HomotopyCategory},
the \emph{derived adjunction}:
\begin{equation}
\kern-6pt\myxymatrix{
{\rm Ho}(\scrD) \ar@<-8pt>[rr]_{\mathbbm{R}}^{\bot} && {\rm Ho}(\scrC) \ar@<-8pt>[ll]_{\mathbbm{L}}
}
\end{equation}
A Quillen adjunction is called a \emph{Quillen equivalence} if this derived adjunction is an equivalence of categories. For instance, simplicial and topological homotopy theory are Quillen equivalent via the singular simplicial complex functor defined in~\eqref{eq:SingularSimplicialComplex},
\begin{equation}\label{QuillenEquivalenceTopsSet}
\kern-6pt\myxymatrix{
\sfTop\ar@{<-}@<+10pt>[rr]^-{|-|}\ar@<-10pt>[rr]_-{\mathrm{Sing}}^{\cong_{\mathrm{Quillen}}}&&\mathsf{sSet}
  }
\end{equation}
Consequently, both represent the homotopy theory of $\infty$-groupoids (`spaces').

\begin{example}{\bf (Classifying spaces)}
\label{exa:ClassifyingSpaces}
Under the Quillen equivalence \eqref{QuillenEquivalenceTopsSet}, the delooping groupoids $\mathbf{B}\sfG$ of a discrete group $\sfG$  from Example \ref{ex:delooping} translate into the \emph{classifying spaces} $\mathsf{BG}\cong|\mathbf{B}\sfG|$, for example the Eilenberg--MacLane space $|\mathbf{B}\mathbb{Z}|\cong K(\mathbb{Z},1)$ (see Example~\ref{exa:DoubleDimRed}). The analogous statement applies to groups with geometric structure, such as locally contractible topological groups, in particular Lie groups. This requires enhancing plain simplicial sets to \emph{simplicial presheaves} over the category of manifolds, hence enhancing plain $\infty$-groupoids to geometric (`cohesive') $\infty$-groupoids, discussed in Section \ref{sec:HigherGeometry} below,
\begin{equation}\label{QuillenEquivalenceTopsSet}
\kern-6pt\myxymatrix{
\sfTop \ar@{<-}@<+10pt>[r]^-{|-|} \ar@<-10pt>[r]_-{\mathrm{Sing}}^-{\cong_{\mathrm{Quillen}}} & \mathsf{sSet}\ar@{<-}@<+10pt>[r]^-{|-|}\ar@<-10pt>[r]_-{\mathrm{Disc}}^-{\bot_{\mathrm{Quillen}}}& \mathsf{Fun}\big(\sfMfd^{\mathrm{op}}, \mathsf{sSet}\big)
}
\end{equation}
The analogous construction then yields classifying spaces $\mathsf{BG}\cong|\mathbf{B}\sfG|$  of topological groups such as the String group $\sfG= \mathsf{String}$, as they appear in Figure \ref{fig:WhiteheadTower}. In fact, any topological group $\sfG$ has a universal principal bundle, that is, a principal $\sfG$-bundle whose total space is contractible. The total space of the universal bundle for $\sfG$ is usually denoted by $\mathsf{EG}$ and its base space, the classifying space for principal $\sfG$-bundles, by $\mathsf{BG}$. Then, any principal $\sfG$-bundle over a topological manifold $X$ is given by the pull-back of $\mathsf{EG}$  by means of a classifying map $\gamma:X\to\mathsf{BG}$, and equivalent principal $\sfG$-bundles over $X$ are described by different representatives of the homotopy classes $[X,\mathsf{BG}]$ of $\gamma$.  This is discussed in detail in
\cite{nikolaus1207}.
\end{example}

\subsection{Higher structures}\label{sec:HigherStructures}

A plain mathematical structure \'a la Bourbaki consists of three types of ingredients:
\smallskip
\begin{enumerate}[label=\roman*),leftmargin=*]
\item a collection of sets;
\item a collection of functions between these;
\item a collection of axiomatic equations between these.
\end{enumerate}
\smallskip
Consider, for example, the definition of a group. We have
\smallskip
\begin{enumerate}[label=\roman*),leftmargin=*]
\item a set $\sfG$ and the one-element set $\ast$;
\item three functions between these sets:
\begin{enumerate}[label=\alph*),leftmargin=*]
\item \emph{multiplication}: a function $(-)\cdot(-):\sfG\times\sfG\to\sfG$;
\item a \emph{neutral} element: a function $e:\ast\to\sfG$;
\item \emph{inverse}: a function $(-)^{-1}:\sfG\to\sfG$;
\end{enumerate}
\item which satisfy the axiomatic equations
\begin{enumerate}[label=\alph*),leftmargin=*]
\item \emph{associativity}: $(g_1\cdot g_2) \cdot g_3 = g_1 \cdot (g_2 \cdot g_3)$;
\item \emph{unitality}: $g \cdot e = e \cdot g=g$;
\item \emph{invertibility}: $g^{-1}\cdot g=e$;
\end{enumerate}
for all $g,g_{1,2,3}\in\sfG$.
\end{enumerate}
\smallskip
A \emph{higher structure} is like a Bourbakian mathematical structures, however, one
\smallskip
\begin{enumerate}[label=\roman*),leftmargin=*]
\item replaces sets by higher homotopy types;
\item replaces functions by homotopies;
\item replaces axiomatic equations by equations which hold up to higher homotopies;
\item enforces \emph{coherence laws}.
\end{enumerate}
\smallskip
Here, \emph{coherence laws} are the required conditions that \emph{iterated (higher) homotopies are unique up to (higher) homotopies}. This process is sometimes also called \emph{(vertical) categorification}.

\begin{figure}[h]
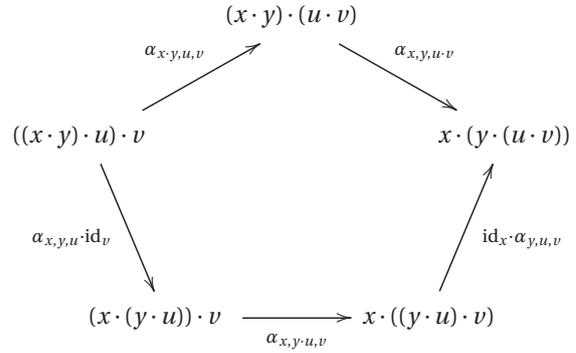

\begin{equation*}
 \hfil\xy
 (0,20)*+{(x\cdot y)\cdot (u\cdot v)}="1";
 (28,4)*+{x\cdot(y\cdot(u\cdot v))}="2";
 (18,-20)*+{ x\cdot((y\cdot u)\cdot v)}="3";
 (-18,-20)*+{(x\cdot(y\cdot u))\cdot v}="4";
 (-28,4)*+{((x\cdot y)\cdot u)\cdot v }="5";
     {\ar^{\alpha_{x,y,u\cdot v}}     "1";"2"}
     {\ar_{\id_x\cdot \alpha_{y,u,v}}  "3";"2"}
     {\ar_{\alpha_{x,y\cdot u,v}}    "4";"3"}
     {\ar_{\alpha_{x,y,u}\cdot \id_v}  "5";"4"}
     {\ar^{\alpha_{x\cdot y,u,v}}    "5";"1"}
\endxy\hfil
\end{equation*}
\vspace{-.7cm}
\caption{The pentagon identity is the higher companion of the associator $\alpha$. It says that the two ways of rebracketing the product of \emph{four} elements in a higher group via the associator must be homotopic via a higher order homotopy. In a 2-group, which has only first-order homotopies, the pentagon identity is thus an actual equality. Its Lie algebraic analogue, satisfied by $L_\infty$-algebras, is shown in Figure~\ref{fig:LiePentagon}.}
\label{fig:IdentityPentagon}
\end{figure}

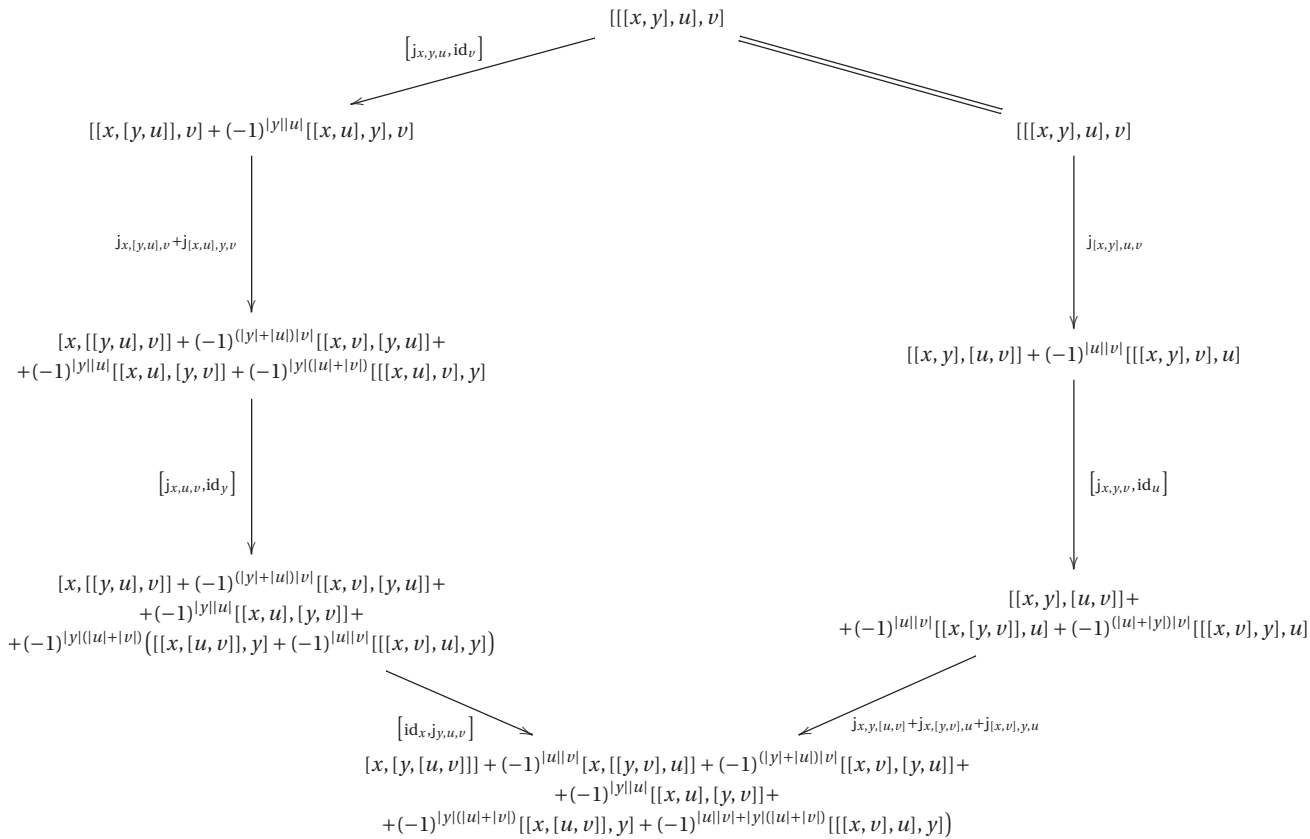
\begin{figure*}[h]
\begin{equation*}
\resizebox{1\hsize}{!}{$
\xymatrix@C=-8em@R=3.2em{
&[[[x,y],u],v] \ar[dl]_{\big[\sfj_{x,y,u},\,\id_v\big]}\ar@{=}[dr]&\\
[[x,[y,u]],v] + (-1)^{|y||u|} [[x,u],y],v]\ar[dd]_{\sfj_{x,[y,u],v}+\sfj_{[x,u],y,v}}&&[[[x,y],u],v]\ar[dd]^{\sfj_{[x,y],u,v}}\\
&&\\
\mbox{$\begin{matrix}[x,[[y,u],v]] + (-1)^{(|y|+|u|)|v|} [[x,v],[y,u]]+\\ + (-1)^{|y||u|}[[x,u],[y,v]] + (-1)^{|y|(|u|+|v|)} [[[x,u],v],y]\end{matrix}$}\ar[dd]_{\big[\sfj_{x,u,v},\id_y\big]} && [[x,y],[u,v]] + (-1)^{|u||v|} [[[x,y],v],u]\ar[dd]^{\big[\sfj_{x,y,v},\,\id_u\big]}\\
&&\\
\mbox{$\begin{matrix}[x,[[y,u],v]] + (-1)^{(|y|+|u|)|v|} [[x,v],[y,u]]+\\ + (-1)^{|y||u|}[[x,u],[y,v]]+\\+ (-1)^{|y|(|u|+|v|)} \big([[x,[u,v]],y] + (-1)^{|u||v|} [[[x,v],u],y]\big)\end{matrix}$}\ar[dr]_{\big[\id_x,\sfj_{y,u,v}\big]}&&\mbox{$\begin{matrix}[[x,y],[u,v]]+\\ +(-1)^{|u||v|}[[x,[y,v]],u] + (-1)^{(|u|+|y|)|v|} [[[x,v],y],u] \end{matrix}$}\ar[dl]^{~~~~~~~~\sfj_{x,y,[u,v]}+\sfj_{x,[y,v],u}+\sfj_{[x,v],y,u}}\\
&\mbox{$\begin{matrix}[x,[y,[u,v]]] + (-1)^{|u||v|} [x,[[y,v],u]]+ (-1)^{(|y|+|u|)|v|} [[x,v],[y,u]]+\\ + (-1)^{|y||u|}[[x,u],[y,v]]+\\+ (-1)^{|y|(|u|+|v|)} [[x,[u,v]],y] + (-1)^{|u||v|+|y|(|u|+|v|)} [[[x,v],u],y]\big)\end{matrix}$}&
}
$}
\end{equation*}
\vspace{-.7cm}
\caption{The higher Lie-algebraic analogue of the pentagon identity from Figure~\ref{fig:IdentityPentagon} relates the different ways to use the Jacobiator in an $L_\infty$-algebra to rebracket four elements. Again, the higher homotopies filling this diagram are trivial in the case of a 2-term $L_\infty$-algebra.}
\label{fig:LiePentagon}
\end{figure*}

\smallskip
\begin{example}
{\bf (2-group)}
\label{exa:2Groups}
The categorification of a group as defined in~\cite{Sinh1975aa,Baez:0307200} is also called a \emph{2-group} and encodes a group structure on a homotopy 1-type. The associativity equality is to be replaced by a choice of homotopy
\begin{equation}
(x \cdot y) \cdot z\ \xrightarrow{~\alpha_{x,y,z}~}\ x \cdot (y \cdot z)
\end{equation}
called the \emph{associator}. The coherence to be imposed on this is the condition that the composite
homotopy
\begin{equation}
((x \cdot y) \cdot u) \cdot v\ \longrightarrow\ x \cdot (y \cdot (u \cdot v))
\end{equation}
is unique in that the two ways of composing associators to achieve this are related by a homotopy-of-homotopies, and, consequently, an equality (by assumption that we have just a 1-type). This coherence condition is called the \emph{pentagon identity} and shown in Figure~\ref{fig:IdentityPentagon}. It implies coherence in the sense that all homotopies of rebracketing any expression, using the given associators, coincide \cite{Lane:1979:415-433}.

Recall the simplicial set $\sfB\sfG$ from Example~\ref{ex:delooping}. For $\sfG$ Abelian, this category carries naturally a 2-group structure with trivial associator.
\end{example}
\smallskip

An ubiquitous higher structures which is particularly relevant to applications in mathematical physics are the higher versions of Lie algebras. To construct these, we regard a Lie algebra as a mathematical structure consisting of
\smallskip
\begin{enumerate}[label=\roman*),leftmargin=*]
\item a vector space $V$ (i.e.~a set with linear structure) over a field $\IK$
\item equipped with a function
\begin{enumerate}[label=\alph*),leftmargin=*]
\item\emph{Lie bracket}: $[-,-]:V \times V \to V$
\end{enumerate}
\item such that the following equations hold:
\begin{enumerate}[label=\alph*),leftmargin=*]
\item \emph{skew-symmetry}: $[x,y] = - [y,x]$:
\item \emph{bilinearity}: $[k_1 x_1 + k_2 x_2, y] = k_1 [x_1, y] + k_2 [x_2,y]$
\item \emph{Jacobi identity}: $[x,[y,z]] = [[x,y], z] + [y,[x,z]]$
\end{enumerate}
for $x,y,z,x_{1,2}\in V$ and $k_{1,2}\in\IK$.
\end{enumerate}
\smallskip
The corresponding \emph{higher structure} is now obtained by \emph{internalising} this in a homotopy theory (an $\infty$-category).
This entails a) promoting the underlying set to a higher homotopy type, b) enhancing the constraining equations by homotopies, and then c) imposing coherence of these homotopies. When interpreted in chain homotopy theory, the resulting higher homotopy Lie algebras are \emph{strong homotopy Lie algebra} (the `strong' refers to coherence) or \emph{$L_\infty$-algebras} (for `L'ie algebras with homotopies up to infinity). This structure was first found in closed string field theory \cite{Zwiebach:1992ie} and then highlighted in \cite{Lada:1992wc}.

We thus arrive at a chain complex $(V_\bullet,\partial)$, endowed with a graded skew-symmetric chain map
\begin{subequations}
\begin{equation}
  [-,-]: V_\bullet \times V_\bullet \to V_\bullet~,
\end{equation}
that is, a graded-skew symmetric bilinear map
\begin{equation}
  [x,y] = -(-1)^{|x||y|} [y,x]
\end{equation}
which respects the differential
\begin{equation}
[\partial x, y] + (-1)^{|x|} [x, \partial y] = \partial[x,y]
\end{equation}
and which satisfies the Jacobi identity up to a specified homotopy called the \emph{Jacobiator}
\begin{equation}
[[x,y],z]\xrightarrow{~\sfj_{x,y,z}~}[x,[y,z]] + (-1)^{|y||z|} [[x,z],y]
\end{equation}
\end{subequations}
for $x,y,z\in V_\bullet$, which, in turn, satisfies higher coherence axioms up to higher homotopies. The Jacobiator coherence condition, which says that the two possible ways of rebracketing four elements are homotopic, is shown in Figure~\ref{fig:LiePentagon}. This is the higher Lie theoretic analogue of the Pentagon identity in Figure~\ref{fig:IdentityPentagon}.

\begin{example} ({\bf Lie 2-algebra}) Let us consider the infinitesimal example of a Lie 2-group, namely a \emph{Lie 2-algebra} or \emph{2-term $L_\infty$-algebra} \cite{Baez:2003aa,Roytenberg:0712.3461}. Here, we have a 2-term chain complex $(V_\bullet,\partial)$ in which only $V_0$ and $V_1$ are non-trivial as well as the chain map $[-,-]:V_\bullet\wedge V_\bullet\rightarrow V_\bullet$ and the Jacobiator.

One can now show that the Jacobiator is encoded in a map $\mu_3:V_0\wedge V_0\wedge V_0\rightarrow V_1$, and together with $\mu_1=\partial$ and $\mu_2$, which is induced by the map $[-,-]$, we obtain the usual higher product or bracket formulation of a 2-term $L_\infty$-algebra, see~\cite{Baez:2003aa}.
\end{example}

\subsection{Higher geometry and classical sigma models}\label{sec:HigherGeometry}

With a general idea of \emph{higher structures} at our disposal, we discuss now general \emph{higher geometric structures}.

Consider any of the categories of affine spaces listed in Table~\ref{tab:IsbellDuality}:
\begin{equation}
\mathsf{Spc}_{\text{affine}}=\begin{cases}
\mathsf{Scheme}_{\text{affine}}\\
\sfMfd\\
\sfSMfd\\
\cdots
\end{cases}.
\end{equation}
We may bootstrap a notion of \emph{generalised spaces}, so-called \emph{$\infty$-stacks}, from these affine spaces.

In particular, the notion of an $\infty$-stack turns out to be the right notion of geometry as seen by classical sigma models. In order to determine a generalised target space $X$, consider for each brane world volume $\Sigma\in\mathsf{Obj}(\mathsf{Spc}_{\text{affine}})$ the collection of sigma model fields $\{\Sigma\to X\}$. For this to be sensible, we need to impose some minimum \emph{consistency conditions}:
\smallskip
\begin{enumerate}[label=\roman*),leftmargin=*]
\item  For every $\Sigma\in\mathsf{Obj}(\mathsf{Spc}_{\text{affine}})$ there should be a simplicial set (see Figure~\ref{fig:IdeaOfSimplicialSets}) of sigma model fields into $X$ and their gauge-of-gauge equivalences, $\Sigma\mapsto\sfMaps(\Sigma,X)$.
\item For every morphism $f\in \mathsf{Mor}(\mathsf{Spc}_{\text{affine}})$ there should be a function pre-composition of sigma model fields with $f$:
\begin{equation}
\kern-6pt\myxymatrix{
\Sigma_1 \ar@{|->}[r]\ar[d]_{f} & \sfMaps(\Sigma_1,X)\\
\Sigma_2 \ar@{|->}[r] & \sfMaps(\Sigma_2,X)\ar[u]_{\text{pre-composition with }f}
}
\end{equation}
\item This should be compatibility with composition of morphisms in $\mathsf{Spc}_{\text{affine}}$.
\end{enumerate}
\smallskip
A moment of reflection shows that these three consistency conditions say nothing but that classical sigma-models see generalised spaces $X$ as probe-assigning functors of the form
\begin{equation}
X:\mathsf{Spc}_{\text{affine}}^{\rm op}\to \mathsf{sSet}~.
\end{equation}
These are also called \emph{simplicial presheaves} on the category of affine spaces.

Hence, the category of generalised spaces modelled on the category of affine spaces should be a full subcategory~\eqref{eq:FullEmbedding} of the functor category
\begin{equation}
\kern-6pt\xymatrix{
\mathsf{GenSpc}  \ar@{^{(}->}[r] & \mathsf{Fun}\big(\mathsf{Spc}_{\text{affine}}^{\rm op},\mathsf{sSet}_{\text{Quillen}}\big)~;
}
\end{equation}
see also Example~\ref{exa:SimplicialHomotopy}. This is also called a \emph{simplicial sheaf topos}
or the \emph{$\infty$-stack $\infty$-topos} over $\mathsf{Spc}_{\text{affine}}$. This is naturally a model category (see Section~\ref{sec:HomotopyTheory}) and thus exhibits the \emph{homotopy theory of higher geometry} modelled on $\mathsf{Spc}_{\text{affine}}$.

The key bootstrap theorem of functorial geometry then says that consistency condition i) becomes true: indeed, in the category of generalised spaces now constructed, we have the following natural identification:
\begin{equation}
\underbrace{\Big\{\sfhom_{\mathsf{Spc}_{\text{affine}}}(-,\Sigma)\Rightarrow X\Big\}}_{\substack{\text{morphisms of}\\ \text{generalised spaces}}}\cong \underbrace{X(\Sigma)\equiv\sfMaps(\Sigma, X)}_{\substack{\text{defining probes}\\ \text{of }X\text{ by }\Sigma}}~.
\end{equation}
This statement is known as the \emph{Yoneda lemma}.

The higher (stacky) geometry thus obtained secretly underlies 
much of gauge field theory 
(see \cite{Shulman:2014:109-126, Benini:2017zjv}) 
and string theory (see \cite{Fiorenza:2013jz, HigherPrequantum}).

\section{Contributions to this volume}

The contributions to this volume can be grouped into the following main topics:
\smallskip
\begin{enumerate}[label=\roman*),leftmargin=*]
 \item Higher differential geometry and higher Lie theory;
 \item Higher structures and M-branes;
 \item Generalised, doubled and exceptional geometry;
 \item Higher structures in classical field theories;
 \item Higher structures and quantum field theories
\end{enumerate}
Below, we shall give a very concise introduction to each of these and list the relevant contributions to this volume.

\subsection{Higher differential geometry and higher Lie theory}

As indicated previously, \emph{higher differential geometry} refers to the formulation of differential geometry within higher geometry. In this formulation, we replace ordinary smooth manifolds with Lie $\infty$-groupoids, which can be regarded as Kan simplicial manifolds. These, in turn, allow us to define higher analogues of all the familiar objects from differential geometry.

The higher analogue of the infinitesimal approximation of a Lie group by a Lie algebra exists, and Lie $\infty$-groupoids differentiate to Lie $\infty$-algebroids, which are most conveniently described in the language of differential graded manifolds. They are crucial in the implementation of the \emph{gauge principle}: together, higher differential geometry and higher Lie theory lead to natural definitions of higher principal bundles with connections.

We have already given a motivational introduction to this topic in the introduction and Section~\ref{sec:ideas}. Let us repeat that classical geometric notions such as manifolds, principal fibre bundles, and Lie groups are often not sufficient to describe structures arising in gauge and string theories. For example, moduli spaces of solutions to certain gauge field equations or brane configurations require the notion of stacks. Roughly speaking, these can be regarded as Lie groupoids or categorified spaces. Likewise, Abelian gerbes, which are central Lie groupoid extensions, and their higher generalisations appear very naturally in string theory. In particular, supergravity form fields and the Kalb--Ramond field of string theory belong to connective structures on the categorified bundles that have (higher) Lie groupoids as their structure groups.

A clean formulation and a resulting clear understanding of these mathematical structures is, in our opinion, a prerequisite for significant progress in M-theory. The following contributions in this volume developed various aspects of higher differential geometry:
\begin{itemize}[leftmargin=20pt]
\item[\cite{contrib:braunackmeyer}] V.~Braunack-Mayer, \emph{Parametrised homotopy theory and gauge enhancement};
\item[\cite{contrib:deser}] A.~Deser, \emph{Pre-NQ manifolds and correspondence spaces: the nilmanifold example};
\item[\cite{contrib:fiorenza}] D.~Fiorenza, H.~Sati, and U.~Schreiber, \emph{The rational higher structure of M-theory};
\item[\cite{contrib:voronov}]T.~T.~Voronov, \emph{Graded geometry, $Q$-manifolds, and microformal geometry};
\item[\cite{contrib:zucchini}] R.~Zucchini, \emph{Wilson surfaces for surface knots.}
\end{itemize}

\subsection{Higher structures and M-branes}

As already mentioned in the introduction, M-theory has already contributed significantly to our understanding of string theory at a heuristic level, but a proper formulation of this theory remains an open problem. One step towards such a formulation is a better understanding of the most fundamental ingredients of M-theory, namely M2- and M5-branes, which are the M-theory analogues of strings and D-branes within string theory.

The interactions of D-branes are mediated by strings ending on them and in the simplest case (e.g.~after `turning off' gravity which amounts to decoupling closed strings), the dynamics of  $N$ coincident flat and parallel D-branes is described by maximally supersymmetric Yang--Mills theory with gauge group $\sU(N)$, dimensionally reduced to the world-volume of the D-branes. An analogue of this description for M2-branes which satisfies most expectations was found a while ago~\cite{Bagger:2007jr,Gustavsson:2007vu,Aharony:2008ug}. However, the corresponding theory for M5-branes, which is often called the \emph{(2,0)-theory} because of its supersymmetry, is still unknown. This theory has attracted a significant amount of interest due to its central role in the web of string theory dualities. There are many indications that the (2,0)-theory is a higher gauge theory and that higher structures are fundamental to its understanding.

Higher structures in the context of M-branes are discussed in the following contributions:
\smallskip
\begin{itemize}[leftmargin=20pt]
\item[\cite{contrib:sorokin}] I.~Bandos, F.~Farakos, S.~Lanza, L.~Martucci, and D.~Sorokin, \emph{Higher forms and membranes in 4D supergravities};
\item[\cite{contrib:chu}] C.-S.~Chu, \emph{Weyl anomaly and vacuum magnetization current of M5-brane in background flux};
\item[\cite{contrib:fiorenza}] D.~Fiorenza, H.~Sati, and U.~Schreiber, \emph{The rational higher structure of M-theory};
\item[\cite{contrib:wolf}] B.~Jur{\v c}o, T.~Macrelli, L.~Raspollini, C.~Saemann, and M.~Wolf, \emph{$L_\infty$-algebras, the BV formalism, and classical fields};
\item[\cite{contrib:huerta}] J.~Huerta, \emph{How space-times emerge from the superpoint};
\item[\cite{contrib:lambert}] N.~Lambert, \emph{M-branes: lessons from M2's and hopes for M5's};
\item[\cite{contrib:saemann}] C.~Saemann, \emph{Higher structures, self-dual strings and 6d superconformal field theories};
\item[\cite{contrib:schmidt}] L.~Schmidt, \emph{Twisted string algebras.}
\end{itemize}

\subsection{Generalised, doubled, and exceptional geometry}

Generalised geometry is, to a certain extent, the study of the geometry of symplectic categorified Lie algebroids, better known as Courant algebroids. These algebroids arise in string theory when studying the geometry underlying T-duality. Let us stress again that this symmetry is one of the key features distinguishing string theory from a theory of point particles. It is also an important ingredient in the web of dualities connecting various string theories.

Formulating a field theory in which T-duality becomes a manifest symmetry was the initial goal of double field theory~\cite{Siegel:1993th,Hull:2007zu}, see also~\cite{contrib:berman}. The corresponding lift to M-theory is known as exceptional field theory. The biggest success of double and exceptional field theory is certainly the construction of appropriate action principles which allow for the expected Kaluza--Klein type reduction to various (super)gravity theories.

A number of aspects of these actions such as their precise mathematical meaning and their global formulations, however, have remained elusive. Clearly, higher geometries and higher symmetry algebras are underlying double and exceptional field theory. Our understanding of these, however, is still in its infancy but crucial to further progress in both fields.

This volume contains the following contributions related to generalised, doubled, and exceptional geometry:
\smallskip
\begin{itemize}[leftmargin=20pt]
\item[\cite{contrib:berman}] D.~S.~Berman, \emph{A Kaluza--Klein approach to double and exceptional field theory};
 \item[\cite{contrib:deser}] A.~Deser, \emph{Pre-NQ manifolds and correspondence spaces: the nilmanifold example};
  \item[\cite{contrib:hohm}] O.~Hohm and H.~Samtleben, \emph{Higher gauge structures  in double and exceptional field theory};
 \item[\cite{contrib:vysoky}] B.~Jur\v co and J.~Vysok\'y, \emph{Effective actions for $\sigma$-models of Poisson--Lie type};
 \item[\cite{contrib:strickland-constable}] C.~Strickland-Constable, \emph{Supergravity fluxes and generalised geometry.}
\end{itemize}

\subsection{Higher structures in classical field theories}

As mentioned in the introduction, the construction of classical string field theory is based on higher algebraic structures: $L_\infty$-algebras in the case of closed string field theory, $A_\infty$-algebras in the case of open string field theory, and a combination of both in the case of open-closed string field theory (with the latter requiring further study).

If we accept the fundamental role of string field theory, it is not too surprising that these homotopy algebras cast a shadow in ordinary classical field theories. This becomes evident in the classical part of the BV formalism applied to a field theory, which is a two-step resolution\footnote{That is, essentially, a reformulation of an interesting space as a cohomology group of a complex.} of the space of classical observables. The latter is obtained from the space of field configurations by dividing out gauge equivalences and subsequently restricting to solutions of the field equations. A Chevalley--Eilenberg resolution (which corresponds to the setup of the ordinary BRST formalism) takes care of the gauge equivalences, while a Koszul--Tate resolution (which introduces the antifields into the formalism) handles the restriction to solutions of the equations of motion. The result is a differential graded vector space which is equivalent to an $L_\infty$-algebra. See \cite{Jurco:2018sby} for a recent discussion of the $L_\infty$-perspective of the BV formalism.

This perspective allows for a nice reformulation of physical concepts in a purely algebraic way. For instance, equivalent field theories have $L_\infty$-algebras which are categorically equivalent.

The contributions related to string field theory and the resulting $L_\infty$-algebras in classical field theories in this volume are:
\smallskip
\begin{itemize}[leftmargin=20pt]
\item[\cite{contrib:grigoriev}] M.~Grigoriev and A.~Kotov, \emph{Gauge PDE and AKSZ-type sigma models}
 \item[\cite{contrib:wolf}] B.~Jur{\v c}o, T.~Macrelli, L.~Raspollini, C.~Saemann, and M.~Wolf, \emph{$L_\infty$-algebras, the BV formalism, and classical fields};
  \item[\cite{contrib:sachs}] I.~Sachs, \emph{Homotopy algebras in string field theory};
 \item[\cite{contrib:kupriyanov}] V.~G.~Kupriyanov,  \emph{$L_\infty$-bootstrap approach to non-commutative gauge theories}.
\end{itemize}

\subsection{Higher structures in quantum field theories}

Eventually, our world is best described by quantum field theories to which the above mentioned classical field theories are mere approximations. First, let us stress again that the BV formalism, which is the relevant formalism for a general quantisation of a classical field theory, is manifestly based on higher structures. This has a number of consequences and implies in particular that higher structures also play a key role in  algebraic quantum field theory and conformal field theory.

In addition, aspects of the correspondence between the equivalence of classical field theories and categorical equivalence of their $L_\infty$-algebras survive quantisation: there are examples of field theories linked by the renormalisation group which are generated by categorically equivalent data.

The contributions to this volume discussing higher structures in the context of quantum field theory are:
\smallskip
\begin{itemize}[leftmargin=20pt]
 \item[\cite{contrib:schenkel}] M.~Benini and A.~Schenkel, \emph{Higher structures in algebraic quantum field theory};
 \item[\cite{contrib:bruinsma}] S.~Bruinsma, \emph{Coloring operads for algebraic field theory};
  \item[\cite{contrib:schweigert}] J.~Fuchs and C.~Schweigert, \emph{Full logarithmic conformal field theory --- an attempt at a status report};
 \item[\cite{contrib:monnier}] S.~Monnier, \emph{A modern point of view on anomalies};
 \item[\cite{contrib:sharpe}] E.~Sharpe, \emph{Categorical equivalence and the renormalization group};
 \item[\cite{contrib:szabo}] R.~J.~Szabo, \emph{Quantization of magnetic Poisson structures.}
\end{itemize}

\bibliography{allbibtex}

\begin{thebibliography}{10}

\bibitem{Witten:1998uk}
E.~Witten,
{\em {Magic, mystery, and matrix},}
\href{http://www.ams.org/notices/199809/witten.pdf}{Not. Amer. Math. Soc. {\bf
  45}  (1998) 1124}.

\bibitem{Duff:1999baa}
M.~J.~Duff,
{\em The world in eleven dimensions,} IoP, Bristol, 1999.

\bibitem{Becker:2007zj}
K.~Becker, M.~Becker, and J.~H.~Schwarz,
{\em {String theory and M-theory: a modern introduction},} Cambridge University
  Press, 2006.

\bibitem{Moore:2014aaa}
G.~W.~Moore,
{\em Physical mathematics and the future,}
talk at \href{http://physics.princeton.edu/strings2014/}{Strings 2014},
  available
  \href{http://www.physics.rutgers.edu/\~gmoore/PhysicalMathematicsAndFuture.pdf}{online}.

\bibitem{Danielsson:2018ztv}
U.~H.~Danielsson and T.~Van~Riet,
{\em {What if string theory has no de Sitter vacua?},}
\href{http://dx.doi.org/10.1142/S0218271818300070}{Int. J. Mod. Phys. D {\bf
  27}  (2018) 1830007} [{\tt
  \href{http://www.arxiv.org/abs/1804.01120}{1804.01120 [hep-th]}}].

\bibitem{Obied:2018sgi}
G.~Obied, H.~Ooguri, L.~Spodyneiko, and C.~Vafa,
{\em {De Sitter space and the swampland},}
{\tt \href{http://www.arxiv.org/abs/1806.08362}{1806.08362 [hep-th]}}.

\bibitem{Akrami:2018ylq}
Y.~Akrami, R.~Kallosh, A.~Linde, and V.~Vardanyan,
{\em {The landscape, the swampland and the era of precision cosmology},}
\href{http://dx.doi.org/10.1002/prop.201800075}{Fortsch. Phys. {\bf 2018}
  (2018) 1800075} [{\tt \href{http://www.arxiv.org/abs/1808.09440}{1808.09440
  [hep-th]}}].

\bibitem{Bunke:2009cba}
U.~Bunke,
{\em {String structures and trivialisations of a Pfaffian line bundle},}
\href{http://dx.doi.org/10.1007/s00220-011-1348-0}{Commun. Math. Phys. {\bf
  307}  (2011) 675}.

\bibitem{Sati:2009ic}
H.~Sati, U.~Schreiber, and J.~Stasheff,
{\em {Differential twisted string and five-brane structures},}
\href{http://dx.doi.org/10.1007/s00220-012-1510-3}{Commun. Math. Phys. {\bf
  315}  (2012) 169} [{\tt \href{http://www.arxiv.org/abs/0910.4001}{0910.4001
  [math.AT]}}].

\bibitem{Sati:2008kz}
H.~Sati, U.~Schreiber, and J.~Stasheff,
{\em {Fivebrane structures},}
\href{http://dx.doi.org/10.1142/S0129055X09003840}{Rev. Math. Phys. {\bf 21}
  (2009) 1197} [{\tt \href{http://www.arxiv.org/abs/0805.0564}{0805.0564
  [math.AT]}}].

\bibitem{Sati:2014}
H.~Sati,
{\em {Nine-brane Structures},}
\href{http://dx.doi.org/10.1142/S0219887815500413}{Int. J. Geom. Meth. Mod.
  Phys. {\bf 12.04}  (2015) 1550041} [{\tt
  \href{http://www.arxiv.org/abs/1405.7686}{1405.7686 [hep-th]}}].

\bibitem{Freed:1999vc}
D.~S.~Freed and E.~Witten,
{\em Anomalies in string theory with D-branes,}
\href{http://dx.doi.org/10.4310/ajm.1999.v3.n4.a6}{Asian J. Math. {\bf 3}
  (1999) 819} [{\tt
  \href{http://www.arxiv.org/abs/hep-th/9907189}{hep-th/9907189}}].

\bibitem{Carey:548736}
A.~L.~Carey, S.~Johnson, and M.~K.~Murray,
{\em {Holonomy on D-Branes},}
available \href{http://cds.cern.ch/record/548736}{online}.

\bibitem{Nikolaus:1207ab}
T.~Nikolaus, U.~Schreiber, and D.~Stevenson,
{\em Principal $\infty$-bundles - general theory,}
\href{http://dx.doi.org/10.1007/s40062-014-0083-6}{J. Homot. Relat. Struct.
  {\bf 10}  (2015) 749} [{\tt
  \href{http://www.arxiv.org/abs/1207.0248}{1207.0248 [math.AT]}}].

\bibitem{Zwiebach:1992ie}
B.~Zwiebach,
{\em {Closed string field theory: quantum action and the BV master equation},}
\href{http://dx.doi.org/10.1016/0550-3213(93)90388-6}{Nucl. Phys. B {\bf 390}
  (1993)~33} [{\tt
  \href{http://www.arxiv.org/abs/hep-th/9206084}{hep-th/9206084}}].

\bibitem{Witten:1985cc}
E.~Witten,
{\em {Non-commutative geometry and string field theory},}
\href{http://dx.doi.org/10.1016/0550-3213(86)90155-0}{Nucl. Phys. B {\bf 268}
  (1986) 253}.

\bibitem{Erler:2013xta}
T.~Erler, S.~Konopka, and I.~Sachs,
{\em {Resolving Witten's superstring field theory},}
\href{http://dx.doi.org/10.1007/JHEP04(2014)150}{JHEP {\bf 1404}  (2014) 150}
  [{\tt \href{http://www.arxiv.org/abs/1312.2948}{1312.2948 [hep-th]}}].

\bibitem{Alexandrov:1995kv}
M.~Alexandrov, A.~Schwarz, O.~Zaboronsky, and M.~Kontsevich,
{\em {The geometry of the master equation and topological quantum field
  theory},}
\href{http://dx.doi.org/10.1142/S0217751X97001031}{Int. J. Mod. Phys. A {\bf
  12}  (1997) 1405} [{\tt
  \href{http://www.arxiv.org/abs/hep-th/9502010}{hep-th/9502010}}].

\bibitem{Baez:1995:6073-6105}
J.~C.~Baez and J.~Dolan,
{\em Higher-dimensional algebra and topological quantum field theory,}
\href{http://dx.doi.org/10.1063/1.531236}{J Math. Phys. {\bf 36}  (1995) 6073}
  [{\tt \href{http://www.arxiv.org/abs/q-alg/9503002}{q-alg/9503002}}].

\bibitem{Basu:2004ed}
A.~Basu and J.~A.~Harvey,
{\em The M2-M5 brane system and a generalized Nahm's equation,}
\href{http://dx.doi.org/10.1016/j.nuclphysb.2005.02.007}{Nucl. Phys. B {\bf
  713}  (2005) 136} [{\tt
  \href{http://www.arxiv.org/abs/hep-th/0412310}{hep-th/0412310}}].

\bibitem{Roytenberg:2002nu}
D.~Roytenberg,
{\em On the structure of graded symplectic supermanifolds and Courant
  algebroids,}
in: ``Quantization, Poisson Brackets and Beyond,'' ed.\ Theodore Voronov,
  Contemp. Math., Vol. 315, Amer. Math. Soc., Providence, RI, 2002
[{\tt \href{http://www.arxiv.org/abs/math.SG/0203110}{math.SG/0203110}}].

\bibitem{Lawvere:1997baa}
F.~W.~Lawvere and S.~H.~Schanuel,
{\em Conceptual mathematics: A first introduction to categories,} Cambridge
  University Press, 1997.

\bibitem{Leinster:1612.09375}
T.~Leinster,
{\em Basic category theory,} Cambridge University Press, 2014
[{\tt \href{http://www.arxiv.org/abs/1612.09375}{1612.09375 [math.CT]}}].

\bibitem{Riehl17}
E.~Riehl,
{\em {Category theory in context},} .

\bibitem{AWODEY:1996:209-237}
S.~Awodey,
{\em Structure in mathematics and logic: A categorical perspective,}
\href{http://dx.doi.org/10.1093/philmat/4.3.209}{Phil. Math. {\bf 4}  (1996)
  209}.

\bibitem{Freyd:1964baa}
P.~Freyd,
{\em Abelian categories - an introduction to the theory of functors,} Harper
  and Row, New York, 1964, available
  \href{https://www.emis.de/journals/TAC/reprints/articles/3/tr3abs.html}{online}.

\bibitem{Schreiber:2018lab}
U.~Schreiber,
{\em Categories and toposes,}
lectures at
  \href{http://www.nesinkoyleri.org/eng/events-detail.php?egitimkod=203}\emph{Modern
  Mathematics Methods in Physics}, Nesin Mathematics Village, June 2018,
  available
  \href{https://ncatlab.org/nlab/show/geometry+of+physics+--+categories+and+toposes}{online}.

\bibitem{Schreiber:2016laa}
U.~Schreiber,
{\em Introduction to homotopy theory,}
lecture notes, Bonn 2016, available
  \href{https://ncatlab.org/nlab/show/Introduction+to+Homotopy+Theory}{online}.

\bibitem{Lambek:1981:111-121}
J.~Lambek,
{\em The influence of Heraclitus on modern mathematics,}
\href{http://dx.doi.org/10.1007/978-94-009-8462-2_6}{Boston Stud. Phil. Sci.
  {\bf 67}  (1981) 111}.

\bibitem{Lawvere:1969aaa}
W.~Lawvere,
{\em Adjointness in foundations,}
\href{https://www.emis.de/journals/TAC/reprints/articles/16/tr16abs.htm}{Dialectica
  {\bf 23}  (1969) 281}.

\bibitem{Kontsevich:9411018}
M.~Kontsevich,
{\em Homological algebra of mirror symmetry,}
{\tt \href{http://www.arxiv.org/abs/alg-geom/9411018}{alg-geom/9411018}}.

\bibitem{Aspinwall:2009isa}
P.~S.~Aspinwall, T.~Bridgeland, A.~Craw, M.~R.~Douglas, A.~Kapustin,
  G.~W.~Moore, M.~Gross, G.~Segal, B.~Szendroi, and P.~M.~H.~Wilson,
{\em {Dirichlet branes and mirror symmetry},} AMS, 2009, available
  \href{http://people.maths.ox.ac.uk/cmi/library/monographs/cmim04c.pdf}{online}.

\bibitem{Kapustin:2006pk}
A.~Kapustin and E.~Witten,
{\em Electric-magnetic duality and the geometric Langlands program,}
\href{http://dx.doi.org/10.4310/cntp.2007.v1.n1.a1}{Commun. Num. Theor. Phys.
  {\bf 1}  (2007)~1} [{\tt
  \href{http://www.arxiv.org/abs/hep-th/0604151}{hep-th/0604151}}].

\bibitem{Frenkel:0906.2747}
E.~Frenkel,
{\em Gauge theory and Langlands duality,}
S{\'e}minaire Bourbaki no 1010, June 2009
[{\tt \href{http://www.arxiv.org/abs/0906.2747}{0906.2747 [math.RT]}}].

\bibitem{KolarMichorSlovak:1993baa}
I.~Kol{\'a}{\v r}, P.~W.~Michor, and J.~Slov{\'a}k,
{\em Natural operations in differential geometry,} Springer, Berlin, 1993,
  available \href{https://www.emis.de/monographs/KSM/}{online}.

\bibitem{Mathai:2003mu}
V.~Mathai and H.~Sati,
{\em Some relations between twisted K-theory and $E_8$ gauge theory,}
\href{http://dx.doi.org/10.1088/1126-6708/2004/03/016}{JHEP {\bf 0403}  (2004)
  016} [{\tt \href{http://www.arxiv.org/abs/hep-th/0312033}{hep-th/0312033}}].

\bibitem{Fiorenza:2016oki}
D.~Fiorenza, H.~Sati, and U.~Schreiber,
{\em {T-duality from super Lie $n$-algebra cocycles for super $p$-branes},}
{\tt \href{http://www.arxiv.org/abs/1611.06536}{1611.06536 [math-ph]}}.

\bibitem{Duff:1987bx}
M.~J.~Duff, P.~S.~Howe, T.~Inami, and K.~S.~Stelle,
{\em {Superstrings in $D = 10$ from supermembranes in $D = 11$},}
\href{http://dx.doi.org/10.1016/0370-2693(87)91323-2}{Phys. Lett. B {\bf 191}
  (1987)~70}.

\bibitem{Braunack-Mayer:2018uyy}
V.~Braunack-Mayer, H.~Sati, and U.~Schreiber,
{\em {Gauge enhancement of super M-branes via parametrized stable homotopy
  theory},}
Comm. Math. Phys. (2018) [{\tt
  \href{http://www.arxiv.org/abs/1806.01115}{1806.01115 [hep-th]}}].

\bibitem{contrib:fiorenza}
D.~Fiorenza, H.~Sati, and U.~Schreiber,
{\em The rational higher structure of M-theory,}
contribution to this volume
[{\tt \href{http://www.arxiv.org/abs/1903.02834}{1903.02834 [hep-th]}}].

\bibitem{Spalinski:1995:73-126}
J.~Spalinski and W.~Dwyer,
{\em Homotopy theories and model categories,}
in: ``Handbook of Algebraic Topology,'' ed. I.~M.~James, p. 73-126, 1995.

\bibitem{UFP2013:aa}
~{The Univalent Foundations Program},
{\em Homotopy type theory - Univalent foundations of mathematics,} Institute
  for Advanced Study, Princeton, 2013, available
  \href{https://homotopytypetheory.org/book/}{online}.

\bibitem{Shulman:1703.03007}
M.~Shulman,
{\em Homotopy type theory: the logic of space,}
in ``New Spaces for Mathematics and Physics,'' eds. G.~Catren, M.~Anel,
[{\tt \href{http://www.arxiv.org/abs/1703.03007}{1703.03007 [math.CT]}}].

\bibitem{Shulman:2014:109-126}
M.~Shulman and U.~Schreiber,
{\em Quantum gauge field theory in cohesive homotopy type theory,}
\href{http://dx.doi.org/10.4204/EPTCS.158.8}{EPTCS {\bf 158}  (2014) 109} [{\tt
  \href{http://www.arxiv.org/abs/1408.0054}{1408.0054 [math-ph]}}].

\bibitem{Quillen:1967bcaa}
D.~Quillen,
{\em Axiomatic homotopy theory,}
in: ``Homotopical algebra,'' Lecture Notes in Mathematics, No. 43, Berlin
  (1967).

\bibitem{Weibel:1994aa}
C.~A.~Weibel,
{\em An introduction to homological algebra,} Cambridge University Press, 1994.

\bibitem{Schreiber:2012lnaa}
U.~Schreiber,
{\em Introduction to homological algebra,}
lectures notes, Utrecht 2012, available
  \href{https://ncatlab.org/schreiber/show/Introduction+to+Homological+Algebra}{online}.

\bibitem{Goerss:1999aa}
P.~Goerss and J.~Jardine,
{\em Simplicial homotopy theory,} Birkh{\"a}user, Boston, 1999.

\bibitem{Bousfield:1976:0-0}
A.~K.~Bousfield and V.~K. A.~M.~Gugenheim,
{\em On PL de Rham theory and rational homotopy type,}
\href{http://dx.doi.org/10.1090/memo/0179}{Mem. AMS {\bf 8}  (1976)}.

\bibitem{Carchedi:1211.6134}
D.~Carchedi and D.~Roytenberg,
{\em On theories of superalgebras of differentiable functions,}
\href{http://www.tac.mta.ca/tac/volumes/28/30/28-30abs.html}{Theor. Appl.
  Categor. {\bf 28}  (2013) 1022} [{\tt
  \href{http://www.arxiv.org/abs/1211.6134}{1211.6134 [math.DG]}}].

\bibitem{Fiorenza:2013nha}
D.~Fiorenza, H.~Sati, and U.~Schreiber,
{\em {Super Lie $n$-algebra extensions, higher WZW models, and super $p$-branes
  with tensor multiplet fields},}
\href{http://dx.doi.org/10.1142/S0219887815500188}{Int. J. Geom. Meth. Mod.
  Phys. {\bf 12}  (2015) 1550018} [{\tt
  \href{http://www.arxiv.org/abs/1308.5264}{1308.5264 [hep-th]}}].

\bibitem{Fiorenza:2016ypo}
D.~Fiorenza, H.~Sati, and U.~Schreiber,
{\em {Rational sphere valued supercocycles in M-theory and type IIA string
  theory},}
\href{http://dx.doi.org/10.1016/j.geomphys.2016.11.024}{J. Geom. Phys. {\bf
  114}  (2017)~91} [{\tt \href{http://www.arxiv.org/abs/1606.03206}{1606.03206
  [hep-th]}}].

\bibitem{Huerta:2018xyh}
J.~Huerta, H.~Sati, and U.~Schreiber,
{\em {Real ADE-equivariant (co)homotopy and super M-branes},}
Comm. Math. Phys. (2018) [{\tt
  \href{http://www.arxiv.org/abs/1805.05987}{1805.05987 [hep-th]}}].

\bibitem{vanNieuwenhuizen:1982zf}
P.~van~Nieuwenhuizen,
{\em {Free graded differential superalgebras},}
in: ``Group Theoretical Methods in Physics,'' Proceedings, 11th International
  Colloquium, Istanbul, Turkey, August 23-28, 1982, available
  \href{https://lib-extopc.kek.jp/preprints/PDF/1983/8302/8302065.pdf}{online}.

\bibitem{D'Auria:1982nx}
R.~D'Auria and P.~Fre,
{\em {Geometric supergravity in $d=11$ and its hidden supergroup},}
\href{http://dx.doi.org/10.1016/0550-3213(82)90376-5}{Nucl. Phys. B {\bf 201}
  (1982) 101}.

\bibitem{Castellani:1991et}
L.~Castellani, R.~D'Auria, and P.~Fre,
{\em {Supergravity and superstrings: a geometric perspective},} World
  Scientific, Singapore, 1991.

\bibitem{Jurco:2016qwv}
B.~Jurco, C.~Saemann, and M.~Wolf,
{\em {Higher groupoid bundles, higher spaces, and self-dual tensor field
  equations},}
\href{http://dx.doi.org/10.1002/prop.201600031}{Fortschr. Phys. {\bf 64}
  (2016) 674} [{\tt \href{http://www.arxiv.org/abs/1604.01639}{1604.01639
  [hep-th]}}].

\bibitem{nikolaus1207}
T.~Nikolaus, U.~Schreiber, and D.~Stevenson,
{\em Principal $\infty$-bundles - presentations,}
\href{http://dx.doi.org/doi:10.1007/s40062-014-0077-4}{J. Homotopy Relat.
  Struct. (2014)} [{\tt \href{http://www.arxiv.org/abs/1207.0249}{1207.0249
  [math.AT]}}].

\bibitem{Sinh1975aa}
H.~X.~Sinh,
{\em Gr-categories,} PhD thesis, Universit{\'e} Paris VII (1975)
available
  \href{https://pnp.mathematik.uni-stuttgart.de/lexmath/kuenzer/sinh.html}{online}.

\bibitem{Baez:0307200}
J.~C.~Baez and A.~D.~Lauda,
{\em Higher-dimensional algebra V: 2-groups,}
\href{http://www.kurims.kyoto-u.ac.jp/EMIS/journals/TAC/volumes/12/14/12-14.pdf}{Theor.
  Appl. Categor. {\bf 12}  (2004) 423} [{\tt
  \href{http://www.arxiv.org/abs/math.QA/0307200}{math.QA/0307200}}].

\bibitem{Lane:1979:415-433}
S.~Mac~Lane,
{\em Natural associativity and commutativity,}
in: ``Saunders Mac Lane Selected Papers,'' p.415-433, 1979.

\bibitem{Lada:1992wc}
T.~Lada and J.~Stasheff,
{\em {Introduction to sh Lie algebras for physicists},}
\href{http://dx.doi.org/10.1007/BF00671791}{Int. J. Theor. Phys. {\bf 32}
  (1993) 1087} [{\tt
  \href{http://www.arxiv.org/abs/hep-th/9209099}{hep-th/9209099}}].

\bibitem{Baez:2003aa}
J.~Baez and A.~S.~Crans,
{\em Higher-dimensional algebra VI: Lie 2-algebras,}
\href{http://tac.mta.ca/tac/volumes/12/15/12-15.pdf}{Theor. Appl. Categor. {\bf
  12}  (2004) 492} [{\tt
  \href{http://www.arxiv.org/abs/math.QA/0307263}{math.QA/0307263}}].

\bibitem{Roytenberg:0712.3461}
D.~Roytenberg,
{\em On weak Lie 2-algebras,}
in: ``XXVI Workshop on Geometrical Methods in Physics 2007,'' ed.\ Piotr
  Kielanowski et al., AIP Conference Proceedings volume 956, American Institute
  of Physics, Melville, NY
[{\tt \href{http://www.arxiv.org/abs/0712.3461}{0712.3461 [math.QA]}}].

\bibitem{Benini:2017zjv}
M.~Benini, A.~Schenkel, and U.~Schreiber,
{\em {The stack of Yang--Mills fields on Lorentzian manifolds},}
\href{http://dx.doi.org/10.1007/s00220-018-3120-1}{Commun. Math. Phys. {\bf
  359}  (2018) 765} [{\tt \href{http://www.arxiv.org/abs/1704.01378}{1704.01378
  [math-ph]}}].

\bibitem{Fiorenza:2013jz}
D.~Fiorenza, H.~Sati, and U.~Schreiber,
{\em {A higher stacky perspective on Chern--Simons theory},}
in ``Proceedings, Winter School in Mathematical Physics: Mathematical Aspects
  of Quantum Field Theory,'' Les Houches, France, January 29-February 3, 2012
[{\tt \href{http://www.arxiv.org/abs/1301.2580}{1301.2580 [hep-th]}}].

\bibitem{HigherPrequantum}
U.~Schreiber,
{\em Higher pre-quantum geometry,}
{\tt \href{http://www.arxiv.org/abs/1601.05956}{1601.05956 [math-ph]}}.

\bibitem{contrib:braunackmeyer}
V.~Braunack-Mayer,
{\em Parametrised homotopy theory and gauge enhancement,}
contribution to this volume
[{\tt \href{http://www.arxiv.org/abs/1903.02862}{1903.02862 [hep-th]}}].

\bibitem{contrib:deser}
A.~Deser,
{\em Pre-NQ manifolds and correspondence spaces: the nilmanifold example,}
contribution to this volume
[{\tt \href{http://www.arxiv.org/abs/1903.02864}{1903.02864 [hep-th]}}].

\bibitem{contrib:voronov}
{\relax Th}.~{\relax Th}.~Voronov,
{\em Graded geometry, $Q$-manifolds, and microformal geometry,}
contribution to this volume
[{\tt \href{http://www.arxiv.org/abs/1903.02884}{1903.02884 [hep-th]}}].

\bibitem{contrib:zucchini}
R.~Zucchini,
{\em Wilson surfaces for surface knots,}
contribution to this volume
[{\tt \href{http://www.arxiv.org/abs/1903.02853}{1903.02853 [hep-th]}}].

\bibitem{Bagger:2007jr}
J.~Bagger and N.~D.~Lambert,
{\em Gauge symmetry and supersymmetry of multiple M2-branes,}
\href{http://dx.doi.org/10.1103/PhysRevD.77.065008}{Phys. Rev. D {\bf 77}
  (2008) 065008} [{\tt \href{http://www.arxiv.org/abs/0711.0955}{0711.0955
  [hep-th]}}].

\bibitem{Gustavsson:2007vu}
A.~Gustavsson,
{\em Algebraic structures on parallel M2-branes,}
\href{http://dx.doi.org/10.1016/j.nuclphysb.2008.11.014}{Nucl. Phys. B {\bf
  811}  (2009)~66} [{\tt \href{http://www.arxiv.org/abs/0709.1260}{0709.1260
  [hep-th]}}].

\bibitem{Aharony:2008ug}
O.~Aharony, O.~Bergman, D.~L.~Jafferis, and J.~M.~Maldacena,
{\em {$\mathcal{N}=6$ superconformal Chern--Simons-matter theories, M2-branes
  and their gravity duals},}
\href{http://dx.doi.org/10.1088/1126-6708/2008/10/091}{JHEP {\bf 0810}  (2008)
  091} [{\tt \href{http://www.arxiv.org/abs/0806.1218}{0806.1218 [hep-th]}}].

\bibitem{contrib:sorokin}
I.~Bandos, F.~Farakos, S.~Lanza, L.~Martucci, and D.~Sorokin,
{\em Higher forms and membranes in 4D supergravities,}
contribution to this volume
[{\tt \href{http://www.arxiv.org/abs/1903.02841}{1903.02841 [hep-th]}}].

\bibitem{contrib:chu}
C.-S.~Chu,
{\em Weyl anomaly and vacuum magnetization current of M5-brane in background
  flux,}
contribution to this volume
[{\tt \href{http://www.arxiv.org/abs/1903.02817}{1903.02817 [hep-th]}}].

\bibitem{contrib:wolf}
B.~Jurco, T.~Macrelli, L.~Raspollini, C.~Saemann, and M.~Wolf,
{\em $L_\infty$-algebras, the BV formalism, and classical fields,}
contribution to this volume
[{\tt \href{http://www.arxiv.org/abs/1903.02887}{1903.02887 [hep-th]}}].

\bibitem{contrib:huerta}
J.~Huerta,
{\em How space-times emerge from the superpoint,}
contribution to this volume
[{\tt \href{http://www.arxiv.org/abs/1903.02822}{1903.02822 [hep-th]}}].

\bibitem{contrib:lambert}
N.~Lambert,
{\em M-branes: lessons from M2's and hopes for M5's,}
contribution to this volume
[{\tt \href{http://www.arxiv.org/abs/1903.02825}{1903.02825 [hep-th]}}].

\bibitem{contrib:saemann}
C.~Saemann,
{\em Higher structures, self-dual strings and 6d superconformal field
  theories,}
contribition to this volume
[{\tt \href{http://www.arxiv.org/abs/1903.02888}{1903.02888 [hep-th]}}].

\bibitem{contrib:schmidt}
L.~Schmidt,
{\em Twisted string algebras,}
contribution to this volume
[{\tt \href{http://www.arxiv.org/abs/1903.02873}{1903.02873 [hep-th]}}].

\bibitem{Siegel:1993th}
W.~Siegel,
{\em Superspace duality in low-energy superstrings,}
\href{http://dx.doi.org/10.1103/PhysRevD.48.2826}{Phys. Rev. D {\bf 48}  (1993)
  2826} [{\tt \href{http://www.arxiv.org/abs/hep-th/9305073}{hep-th/9305073}}].

\bibitem{Hull:2007zu}
C.~M.~Hull,
{\em {Generalised geometry for M-theory},}
\href{http://dx.doi.org/10.1088/1126-6708/2007/07/079}{JHEP {\bf 0707}  (2007)
  079} [{\tt \href{http://www.arxiv.org/abs/hep-th/0701203}{hep-th/0701203}}].

\bibitem{contrib:berman}
D.~S.~Berman,
{\em A Kaluza--Klein approach to double and exceptional field theory,}
contribution to this volume
[{\tt \href{http://www.arxiv.org/abs/1903.02860}{1903.02860 [hep-th]}}].

\bibitem{contrib:hohm}
O.~Hohm and H.~Samtleben,
{\em Higher gauge structures in double and exceptional field theory,}
contribution to this volume
[{\tt \href{http://www.arxiv.org/abs/1903.02821}{1903.02821 [hep-th]}}].

\bibitem{contrib:vysoky}
B.~Jurco and J.~Vysoky,
{\em Effective actions for $\sigma$-models of Poisson--Lie type,}
contribution to this volume
[{\tt \href{http://www.arxiv.org/abs/1903.02848}{1903.02848 [hep-th]}}].

\bibitem{contrib:strickland-constable}
C.~Strickland-Constable,
{\em Supergravity fluxes and generalised geometry,}
contribution to this volume
[{\tt \href{http://www.arxiv.org/abs/1903.02842}{1903.02842 [hep-th]}}].

\bibitem{Jurco:2018sby}
B.~Jurco, L.~Raspollini, C.~Saemann, and M.~Wolf,
{\em {$L_\infty$-algebras of classical field theories and the
  Batalin--Vilkovisky formalism},}
{\tt \href{http://www.arxiv.org/abs/1809.09899}{1809.09899 [hep-th]}}.

\bibitem{contrib:grigoriev}
M.~Grigoriev and A.~Kotov,
{\em Gauge PDE and AKSZ-type sigma models,}
contribution to this volume
[{\tt \href{http://www.arxiv.org/abs/1903.02820}{1903.02820 [hep-th]}}].

\bibitem{contrib:sachs}
I.~Sachs,
{\em Homotopy algebras in string field theory,}
contribution to this volume
[{\tt \href{http://www.arxiv.org/abs/1903.02870}{1903.02870 [hep-th]}}].

\bibitem{contrib:kupriyanov}
V.~G.~Kupriyanov,
{\em $L_\infty$-bootstrap approach to non-commutative gauge theories,}
contribution to this volume
[{\tt \href{http://www.arxiv.org/abs/1903.02867}{1903.02867 [hep-th]}}].

\bibitem{contrib:schenkel}
M.~Benini and A.~Schenkel,
{\em Higher structures in algebraic quantum field theory,}
contribution to this volume
[{\tt \href{http://www.arxiv.org/abs/1903.02878}{1903.02878 [hep-th]}}].

\bibitem{contrib:bruinsma}
S.~Bruinsma,
{\em Coloring operads for algebraic field theory,}
contribution to this volume
[{\tt \href{http://www.arxiv.org/abs/1903.02863}{1903.02863 [hep-th]}}].

\bibitem{contrib:schweigert}
J.~Fuchs and C.~Schweigert,
{\em Full logarithmic conformal field theory --- an attempt at a status
  report,}
contribution to this volume
[{\tt \href{http://www.arxiv.org/abs/1903.02838}{1903.02838 [hep-th]}}].

\bibitem{contrib:monnier}
S.~Monnier,
{\em A modern point of view on anomalies,}
contribution to this volume
[{\tt \href{http://www.arxiv.org/abs/1903.02828}{1903.02828 [hep-th]}}].

\bibitem{contrib:sharpe}
E.~Sharpe,
{\em Categorical equivalence and the renormalization group,}
contribution to this volume
[{\tt \href{http://www.arxiv.org/abs/1903.02880}{1903.02880 [hep-th]}}].

\bibitem{contrib:szabo}
R.~J.~Szabo,
{\em Quantization of magnetic Poisson structures,}
contribution to this volume
[{\tt \href{http://www.arxiv.org/abs/1903.02845}{1903.02845 [hep-th]}}].

\end{thebibliography}
\bibliographystyle{prop2015}

\end{document}